\documentclass[a4paper,11pt]{article}
\pdfoutput=1 

\usepackage{jheppub} 

\usepackage[T1]{fontenc}
\usepackage[utf8]{inputenc}						
\usepackage{lmodern}
\usepackage{textcomp}
\usepackage{microtype}
\usepackage[inline]{enumitem}
\usepackage{scalerel}
\usepackage{subdepth}
\usepackage{ulem}
\usepackage{subfiles}
\usepackage{import}
\usepackage{appendix}

\usepackage[american]{babel}
\usepackage[style=english]{csquotes}
\usepackage{alphabeta}
\usepackage{upgreek}
\usepackage[all,USenglish]{foreign}
\usepackage{xcolor}
\usepackage{verbatim}

\usepackage[subrefformat = parens]{subcaption}
\usepackage{graphicx}
\usepackage{tabularx}							

\usepackage{mathtools}
\usepackage{amssymb}
\usepackage{dsfont}
\usepackage{physics}						
\usepackage{slashed}						
\usepackage{units}
\usepackage{bm}

\allowdisplaybreaks[1]

\usepackage{tikz}
\usepackage{tikzscale}
\usepackage{tikz-feynman}

\usepackage{hyperref}
\hypersetup{
  hidelinks,
  colorlinks=true,
  linkcolor={black},
  citecolor={blue!50!black},
  linktoc=all,
  pdfencoding=auto,
  psdextra,
  breaklinks=true,
  pdfcreator={},
  pdfproducer={}
}
\usepackage[capitalize,nameinlink]{cleveref}

\graphicspath{figures/} 

\newcommand{\rmd}{\mathrm{d}}
\newcommand{\rme}{\mathrm{e}}
\newcommand{\rmi}{\mathrm{i}}

\newcommand{\pp}{\boldsymbol{p}}
\newcommand{\kp}{\boldsymbol{k}}
\newcommand{\lp}{\boldsymbol{l}}
\newcommand{\qp}{\boldsymbol{q}}

\newcommand{\xv}{\boldsymbol{x}}
\newcommand{\yv}{\boldsymbol{y}}

\newcommand{\zv}{\boldsymbol{z}}
\newcommand{\Pp}{\boldsymbol{P}}

\newcommand{\as}{\alpha_{S}}
\newcommand{\NC}{N_{\mathrm{c}}}
\newcommand{\calO}{\mathcal{O}}
\newcommand{\Dipper}{\textsc{McDipper}}

\newcommand{\pT}{p_\perp}

\newcommand{\qtil}{\tilde{\qp}}
\newcommand{\ktil}{\tilde{\kp}}

\newcommand{\calH}{\mathcal{H}}

\newcommand{\calD}{\mathcal{D}}

\newcommand{\bara}{\bar{a}}
\newcommand{\barb}{\bar{b}}

\newcommand{\bk}{\bar{\kp}}

\newcommand{\bvp}{\bar{v}^+}
\newcommand{\bvm}{\bar{v}^-}

\newcommand{\PdPbar}{\Pp\cdot\bar{\Pp}}
\newcommand{\PXPbar}{\Pp\!\times\!\bar{\Pp}}

\definecolor{oscarC}{RGB}{22, 156, 172}

\newcommand{\ktfact}{$k_\perp$-factorization}

\title{Initial momentum anisotropies in the kT-factorization of the CGC I: Gradient Expansion}

\author[a]{Oscar Garcia-Montero}

\affiliation[a]{ Instituto Galego de Física de Altas Enerxías (IGFAE), Universidade de Santiago de Compostela\\ 
	E-15782 Galicia, Spain}

\emailAdd{oscarjesus.garcia@usc.es}

\abstract{
We revisit single-inclusive gluon production in the dilute--dilute limit of the Color Glass Condensate and show that $k_\perp$-factorization is only the zeroth order of a systematic gradient expansion in the transverse positions of the two colliding sources, organized in powers of $1/(Q_sR)$, with $Q_s$ the saturation scale and $R$ the size over which the sources vary. Collecting terms order by order builds a ladder of scalar and tensor structures from the local distributions and their transverse derivatives, yielding an extended $k_\perp$-factorized formula that retains the spatial dependence of the sources and, through kinetic moments of the spectrum, the energy-momentum tensor. Repeating the expansion from a real-time computation of the classical Glasma fields in the future light cone reproduces the momentum-space result in the eikonal limit, but additionally retains coherent interference between amplitude and conjugate. This adds terms to the ladder (flow responses, chromo-electric--magnetic interference, and a longitudinal energy flux) further suppressed by powers of the large outgoing momentum $|\pp|\gg Q_s$.
The hierarchy thus reveals momentum-space anisotropies present in the initial state before any hydrodynamic evolution, which leading $k_\perp$-factorization misses by construction. Since each order is a local operator on the same unintegrated distributions, these corrections can be added directly to existing saturation-based initial-state models, giving a dynamically generated initial momentum anisotropy and a more faithful early-time energy-momentum tensor without abandoning the tractable factorized form. Set by gradients of the transverse density profiles, they grow towards more dilute, fluctuation-dominated systems, e.g. light-ion collisions. Phenomenological consequences will be developed in a companion paper.
}

\begin{document} 

\maketitle

\flushbottom

\section{Introduction}
\label{sec:intro}

When heavy-ions are collided at ultrarelativistic energies at the Large Hadron Collider (LHC) and in the Relativistic Heavy Ion Collider (RHIC), the quarks and gluons inside the nuclei interact to create a medium of deconfined hot nuclear matter, the Quark-Gluon Plasma (QGP)
\cite{Elfner:2022iae}. The produced matter, is, to a remarkable degree, described by relativistic viscous hydrodynamics  ~\cite{Heinz:2013th,Busza:2018rrf}. 
Because of this significant progress has been made in understanding the QGP's nearly perfect fluid behavior, leading generally to complex stage-by-stage descriptions where QGP transitions from a short non-equilibrium stage, followed by viscous hydrodynamics, and finally to hadronic transport models. Nevertheless, the predictions for bulk observables and multiparticle correlations produced by these frameworks are sensitive to the conditions handed over at the earliest
times of the collision, where spatial anisotropies in the incoming nuclei imprint correlations onto the early QGP~\cite{Nijs:2020roc,JETSCAPE:2020mzn}.
Understanding the physical mechanisms that create these structures is fundamental to reproduce the experimental data accurately, but also to reduce the systematic uncertainties clouding our community's effort to extract the fundamental properties of QGP (and hence of QCD) from HIC experimental data.

 A broad range of models have been created to serve as initial conditions to the ensuing hydrodynamical evolution. On one hand, there are parametric, Glauber model-inspired  prescriptions such as TRENTo~\cite{Moreland:2014oya}, which deposit entropy or energy through a flexible ansatz built from the nuclear thickness functions and are then calibrated to data in global Bayesian analyses~\cite{Bernhard:2019bmu,Nijs:2020roc,JETSCAPE:2020mzn}. Additionally, there are models based on high-$x$ production of partons, such as the (MC-) EKRT framework~\cite{Eskola:1999fc,Kuha:2024kmq}, which builds the deposited energy from perturbative minijet production from pQCD, which allows it to include effectively event-by-event fluctuations.
 
On the other hand, evolution of the initial state can be dynamically modeled using the Color Glass Condensate (CGC), an effective description of Quantum Chromodynamics (QCD)~\cite{McLerran:1993ni,McLerran:1993ka,McLerran:1994vd}, where the small-
$x$ gluons of the incoming nuclei are described by classical color fields sourced by the large-$x$ color sources (hard partons including valence quarks). To perform this evolution, the standard procedure for the boost-invariant initial conditions is to solve the classical Yang–Mills (CYM) equations in real time, such as in the phenomenologically successful IP-Glasma~\cite{Schenke:2012wb,Schenke:2012hg}. 
Additionally, an important feature of this evolution, is that it propagates the initial chromoelectric and chromomagnetic fields, producing, through interactions,  momentum-space anisotropy already at the initial stage, before any hydrodynamic response. Nevertheless, extending such real-time simulations to the full  rapidity-dependent case is not only computationally demanding \cite{Schenke:2022mjv,Schlichting:2020wrv}, but also it is not yet fully clear how to relate the longitudinal structure of the color currents to the small-$x$ structure
of the colliding nuclei~\cite{Schenke:2017rkc, McDonald:2020oyf}.
A more tractable alternative is to work in the dilute–dilute (high-momentum) limit, where single-inclusive gluon production reduces to the 
{\ktfact} ansatz of the CGC, which is  the form computed in real time in \cite{Ipp:2021lwz,Ipp:2024ykh}, and  employed by models such as the  \Dipper{}~\cite{Garcia-Montero:2023gex,garcia_montero_2026_20179552, Garcia-Montero:2025bpn}, which can be viewed as the  IP-Glasma restricted to this factorized regime and integrated up to $Q_s \tau \gtrsim1 $, where $Q_s$ corresponds to the effective saturation scale.  While in the \Dipper{} one recovers a clear connection between the low-$x$ structure of the nuclei and the produced particles, the factorized formula evaluates the sources locally and no longer resolves the field dynamics that generate the anisotropy, so at leading order the deposited transverse momentum is isotropic. By keeping the $x$-dependence of the uGDFs, the rapidity dependence of the corrections is retained, and thus the gradient tower correcting it inherits rapidity dependence.

Except for the IP-Glasma, all of the frameworks mentioned above share, the same general feature, in which the conversion of the spatial geometry of the collision into a momentum-space anisotropy is delegated almost entirely to the hydrodynamic response~\cite{Lappi:2006xc,Borghini:2022iym}. What remains far less settled is the momentum structure of the state handed over to hydrodynamics, i.e. how much of that anisotropy is already set at the level of the produced gluons within the structure of the energy-momentum tensor~\cite{Chen:2015wia,Ipp:2021lwz,Ipp:2024ykh}, the basic piece to construct the hydrodynamical evolution. It is important to note that this momentum structure is degenerate with the shear and bulk viscosities extracted from flow data in global fits~\cite{Gardim:2014tya,Liyanage:2023nds}, so any anisotropy produced before the hydrodynamic stage is effectively reabsorbed into the fitted transport coefficients.

In this work we tackle the question of how the momentum anisotropies are  handed over to hydrodynamics in the dilute-dilute limit of the CGC. We show that the standard $k_\perp$-factorized result is only the leading term of a systematic gradient expansion of the dilute–dilute production formula in the transverse positions of the two incoming sources, organized in powers of $1/(Q_sR)$, where $R$ is the scale at which the source vary in transverse space. We 
 approach this in two complementary ways. First, we compute single-inclusive gluon production in the traditional momentum-space realization~\cite{Blaizot:2004wu,Dumitru:2001ux}. Since the construction of the energy-momentum tensor from gluon on-shell yields lacks, by construction, infrared, field-like dynamics, we confirm our findings through an independent real-time calculation of $\expval{T^{\mu\nu}}$ from the classical Glasma fields. In these expansions,  each additional gradient acts as a transverse derivative of the local unintegrated distributions, yielding  order-by-order  terms which keep the factorization of the UGDs. Our main result is that the two routes agree where they should, the limit in which the large momentum field modes behave like kinematic on-shell plane waves. However,  the field computation carries more information, retaining the coherent interference between the field modes and their conjugates. The result is that these infrared corrections give longitudinal energy and momentum flux, $T^{\tau\eta}$, $T^{\eta i}$, which strictly vanish in the yield computation. In this way, we have not only quantified the terms which lead to transverse anisotropies, e.g. $T^{xx}-T^{yy}\neq0$ and a shear $T^{xy}$, but also identified the mechanism which creates non-vanishing transverse momentum flux.
 
We would like to note that results like these, including longitudinal flux, were obtained earlier in Refs.~\cite{Chen:2013ksa,Chen:2015wia} in a small-$\tau$ expansion of the Glasma. On one hand, this work's results were derived using a large-momentum treatment, rendering the domain of validity of the approximation schemes differently. Additionally, the physical origin of the gradients in Chen et al. was the breaking of translation invariance within the incoming nuclei, which results in the breaking of the factorization of the momentum and impact parameter dependence of the uGDFs
\footnote{This can be thought as a large source-size example of a result like in  Ref.~\cite{Garcia-Montero:2025ekv}, where we observed a breaking of translation invariance when computing the dipole two-point function of a gaussian color distribution, in the context of Deeply Inelastic Scattering (DIS). In this case, its dependence on the impact parameter no longer factorized from that on the relative transverse separation, meaning that now not only the size, but also the orientation of the incoming dipole coupled to the profile itself. One can expand around homogeneity, and obtain tensor structures now within the gluon distribution itself.}, 
 while our anisotropies originate from the real-time interactions within the collisions. The difference stems from the fact that in our computation, the produced gluons are resolving spatially the incoming sources through interactions. It is both the non-triviality of the profiles, in combination to resulting imbalance of the profiles arising for, e.g. non central collisions, what gives way to these gradient corrections. In principle, a similar expansion may be carried directly expanding the uGDFs, presumably adding to the effect. In principle, one could carry out an analogous gradient expansion of the uGDs themselves, producing a second tower of corrections on top of the one derived here. However, unlike the results in this work, such an expansion would depend on the model chosen for the uGDFs.
 
 It is important to note that this  mechanism of correlation creation is  different from earlier works \cite{Dumitru:2008wn,Dusling:2012iga,Mace:2018vwq,Lappi:2015vta,Schlichting:2016sqo} where $n$-gluon correlations where generated via local $n$-point correlators. In contrast, our anisotropies arise from collective dynamics, just non-hydrodynamical collective behavior. A consequence of this is that anisotropies produced in this work can be already present in the single gluon spectrum, even when looking at smooth nuclear profiles, as long as there is a non-vanishing impact parameter. We have here then a dynamical mechanism translates long-range geometry into the early-times microscopic gluon physics.

Finally, we would like to state that a useful feature of the formulation in this work is that each gradient correction remains a local operator acting on the same unintegrated distributions, so it can be incorporated into saturation-based initial-state models like the \Dipper{} to generate realistic event-by-event profiles, giving them dynamically generated initial momentum anisotropies. Because the effect is controlled by density
gradients, it is expected to be most pronounced in the fluctuation-dominated
environments of light-ion collisions such as O+O and
Ne+Ne~\cite{Garcia-Montero:2026oonene}, where it may bear on the question of how,
and how quickly, such systems approach a hydrodynamic
regime~\cite{Garcia-Montero:2026Salgado}. To get intuition on how these extra terms affect the initial state we accompany  this work with a more phenomenologically inclined paper \cite{Garcia-Montero:2026NextPaper}, where the production formulas will be computed fully analytically for the simple Golec-Biernat-Wüsthoff Gaussian model~\cite{Golec-Biernat:1999qor,Kowalski:2003hm,Kowalski:2006hc}, which will allow us to express them in terms of the gradients of the GBW saturation scale $Q^2_{A,B}(x,\xv)$.

The paper is organized as follows. In \cref{sec:tmunu} we review single-inclusive
gluon production in the dilute--dilute limit of the CGC and carry out the
gradient expansion of the production formula, deriving the extended
$k_\perp$-factorized result order by order. We then translate the gluon spectrum
into the energy-momentum tensor and work out corrections up to fourth order in the gradients. In \cref{sec:real_time} we perform an analogous  expansion, evaluating directly the energy-momentum tensor from its corresponding field definition and evolution in real time. From this expansion we recover the results from \cref{sec:tmunu}, and additional terms which effectively represent an interpolation between the particle-like UV dynamics and the IR-dominated field-like dynamics. We summarize the results and present conclusions and outlook in \cref{sec:summary}. Finally, in \cref{app:tmunu-reduction} an analytical reduction of the integration kernel presented in \cref{sec:real_time} is presented for two connected physical limits. 

\paragraph{Notation and coordinates:} In this work, bold symbols  correspond to  transverse positions, $\xv, \yv$, and momenta, $\kp,\qp$. Their norm in transverse 2D space will be denoted as, e.g. $k_\perp$, since non-bold versions of vectors correspond to their 4-vector version. For example $q=(q^+,q^-, \qp)$. We will use a mostly minus metric, signature $(+,-,-,-)$, and use light-cone coordinates in the Kogut-Soper convention~\cite{Brodsky:1997de}, meaning $x^\pm=(t\pm z)/\sqrt{2}$. Additionally, we will use Milne coordinates where we can express $(t,z)$ by the proper time $\tau$ and the spacetime rapidity $\eta$, $t = \tau\cosh\eta$ and $z = \tau\sinh\eta,$ with inverse $\tau = \sqrt{t^2 - z^2}$ and $\eta = \operatorname{arctanh}\!\Big(\tfrac{z}{t}\Big) = \tfrac12\ln\frac{t+z}{t-z}=\tfrac12\ln\frac{x^+}{x^-}$. These coordinates are restricted the inside of the light cone. Finally, for both Milne and light cone coordinates, the transverse coordinates are equivalent to the Cartesian ones. 

\section{The energy-momentum tensor from particle production in the CGC}
\label{sec:tmunu}

The aim of this section is to extend the simplified energy-density formula of
$k_\perp$-factorization to the complete $T^{\mu\nu}$ of the gluons produced in a
dilute--dilute collision. We will first fix the CGC objects and kinematics
(\cref{sec:cgc-setup}), and define the energy-momentum tensor, 
$T^{\mu\nu}$, of a collection of particles (\cref{sec:tmunu-spectrum}). For this, we will use the spectrum of gluons obtained via single-inclusive gluon production channel (\cref{sec:gradient-expansion}). This formula, simplified in the past for infinite nuclei, will be improved by setting a systematic gradient expansion on the transverse-spatial gradients of the incoming unintegrated gluon distributions (uGDFs), in  \cref{sec:gradient-expansion}. This connection to the uGDFs will allow the production of initial state anisotropies on the gluon spectrum, creating an initial shear tensor. 
\subsection{The dilute-dilute limit of the CGC}
\label{sec:cgc-setup}

Consider the collision of two ultrarelativistic nuclei, $A$, moving along the positive light cone $t=z$ and carrying a large light-cone momentum $P^+$, and $B$, moving along the negative light cone with large $P^-$. In
the CGC each nucleus is described as a semi-classical wave packet of small-$x$ gluons radiated by its fast, large-$x$ partons, which act as a classical color-charge density. For a more comprehensive review of the state-of-the-art of the CGC, as open questions, we refer the reader to Ref.~\cite{Garcia-Montero:2025hys}. The field generated by the right-moving target  $A$ can be found from the large-$x$ color sources directly from the  classical Yang-Mills (CYM), $\mathcal{D}_\mu \mathcal{F}_A^{\mu\nu}= \mathcal{J}_A^\nu\,$ where $\mathcal{J}_A^\nu\, = \delta^{-\nu} \rho_{A,a} t^a$. By solving in covariant gauge, $\partial_\mu \mathcal{A}_A^\mu = 0$, we get 
\begin{equation}
	\mathcal{A}_A^-(x^+, \xv)
	= g\int_{\zv} G(\xv-\zv) \rho_A(x^+,\zv) 
	\label{eq:BG_fields}
\end{equation}
where $G$ stands for the gluon transverse propagator,
\begin{equation}
	G(\xv-\zv)  = \int_{\kp} \frac{\rme^{-\rmi \kp\cdot(\xv-\zv)}}{\kp^2+m^2}
\end{equation}
where the parameter $m^2$ is introduced as an infrared regulator for the field response. The produced field is then linear in the source and is given by 
\begin{equation}
	\begin{split}
		\mathcal{A}_A^-(x^+, \xv)
		&= g\, G(\kp)\tilde{\rho}_A(x^+,\kp)
	\end{split}
	\label{eq:BG_fields_FT}
\end{equation}
in transverse momentum space, with $G(\kp)=(\kp^2+m^2)^{-1}$.

Starting from the moment of collision and onwards, the evolution of the created medium can be modeled by solving the CYM equations for the joint sources of both incoming nuclei.
Then, in the positive lightcone, the
\emph{dilute} (or weak-field) limit consists of solving the field equations
perturbatively in the two color-charge densities and keeping the lowest
non-trivial order in \emph{each}, so that $a^\mu=\calO(\rho_A\rho_B)$ and the
produced gluon results from a single exchange with each nucleus. What is important here is that this power-counting holds when
$g\rho/\kp^2\ll1$, i.e. when the characteristic transverse momenta probed within the process of interest lies above the saturation scale,
$p_\perp\gtrsim Q_s$ (or
equivalently, for sufficiently small sources). The higher orders in
$\rho$ that resum into Wilson lines in the dense-target regime are not included in  this limit. 

Physical observables follow from averaging over the color-charge configurations
of the two nuclei, denoted $\langle\,\cdot\,\rangle$, which at the Gaussian
(McLerran--Venugopalan) level is fixed entirely by the two-point functions
$\langle\rho^\dagger\rho\rangle$ introduced below. Throughout, $g$ is the QCD
coupling ($\as=g^2/4\pi$) and $\NC$, $C_F$ the usual color factors.

\subsection{The energy-momentum tensor from single gluon production}
\label{sec:tmunu-spectrum}

The goal of this section is to build intuition on how to  extend the simplified energy density formula for the complete $T^{\mu\nu}$ arising from single gluon production in the limit of \ktfact.  In this way, for a collection of particles in set  of general coordinates, one can compute 
\begin{equation}
	T^{\mu\nu} =\frac{\nu}{(2\pi)^3} \int \frac{\rmd ^3 p } {p^0} \sqrt{-g} p^{\mu}p^{\nu} f(x,p) 
\end{equation} 
where the momentum in Milne coordinates can be expressed as 
$p^{\tau} = \pT\cosh(y-\eta)$ and 
	$p^{\eta} = \frac{\pT}{\tau}\sinh(y-\eta)$
since we are dealing with massless gluons. The transverse momentum  $\pp$ remains the same as in Cartesian. Here, $y$ stands for the  momentum-space rapidity. In this frame, our variables depend on the boost-invariant difference $y-\eta$. Using the transformation leads then to our Milne-coordinates energy-momentum tensor
\begin{equation}
	T^{\mu\nu} = \frac{\nu}{(2\pi)^3}\int \rmd ^2 \pp \,\rmd y\,  p^{\mu}p^{\nu} f(x,p) 
\end{equation} 

We now specialize to gluon production in the \ktfact{} limit, which contains the bulk of the physics we are interested in. In this limit of the CGC, as implemented in the \Dipper{}~\cite{Garcia-Montero:2025hys}, the relevant object is ${dN}/{d^2\xv\,d^2\pp\,dy}$. At LO in the CGC it is known that~\cite{Garcia-Montero:2023gex}
\begin{equation}
	\frac{dN}{d^2\xv d\eta d^2\pp dy} = \frac{dN}{d^2\xv d^2\pp dy}\, \delta (\eta -y)\,.
\end{equation}

We can trace a fixed proper time hypersurface to find the  particle distribution, namely
\begin{equation}
	f(x,p) = \frac{(2\pi)^3}{\nu} \frac{1}{\tau p^\tau}\frac{dN}{d^2\xv d^2\pp dy}  \delta (\eta -y)\,, 
\end{equation}
to finally compute the kinetic energy-momentum tensor,
\begin{equation}
	\tau T^{\mu\nu} = \int \rmd ^2 \pp \,\rmd y\,    \frac{p^{\mu}p^{\nu} }{ p^\tau}\frac{dN}{d^2\xv d^2\pp dy}  \delta (\eta -y)
	\label{eq:Tmunu}
\end{equation} 

Due to the $\eta-y$ dependence in Milne coordinates as well as the appearance of $\delta(\eta - y)$, all the $\eta$ components of $T^{\mu\nu}$ 
vanish.

\subsection{Single gluon production in the dilute-dilute limit of the CGC}
\label{sec:gradient-expansion}

The single-inclusive gluon spectrum can be defined as the Lorentz-invariant number of gluons
produced per unit phase space. This can be  written in terms of the modulus squared of the
production amplitude $\mathcal{M}_g$ as

\begin{equation}
\frac{\rmd N_g}{\rmd^2 \pp\,\rmd y }
	= \frac{g^2}{2(2\pi)^3} |\mathcal{M}_g|^2
\end{equation}

The amplitude, yield and cross sections were computed previously in the dilute-dilute limit in a diverse set of gauges and methods, see  \cite{Blaizot:2004wu, Dumitru:2001ux,Lappi:2003bi,Kovner:1995ja, Kovner:1995ts,Kovchegov:1997ke}. We will use here the momentum space covariant gauge computation \cite{Blaizot:2004wu}. With the conventions stated in \cref{sec:cgc-setup}, we can write the gluon yield as 

\begin{equation}
	\begin{split}
\frac{\rmd N_g}{\rmd^2 \pp\,\rmd y }
		= \frac{g^2}{4(2\pi)^3 C_F} &
		\int_{\boldsymbol{k}_1 \bar{\boldsymbol{k}}_1,\boldsymbol{k}_2 \bar{\boldsymbol{k}}_2}	\Pi_2(\{q_i\})
		h\left(\boldsymbol{k}_1,\boldsymbol{k}_2,\bar{\boldsymbol{k}}_1,\bar{\boldsymbol{k}}_2\right)\\
			&\times 
		\frac{g^2 \left\langle \rho^\dagger_{A,a}(\boldsymbol{k}_1)
		\rho_{A,a}(\bar{\boldsymbol{k}}_1)\right\rangle}{\kp_1^2\,\bk_1^2}		
		\frac{g^2\left\langle  \rho^\dagger_{B,b}(\boldsymbol{k}_2)\,\rho_{B,b}(\bar{\boldsymbol{k}}_2)	\right\rangle}{\kp_2^2\, \bk_2^2}
	\end{split}
	\label{eq:yieldsr}
\end{equation}
Here, for notational simplicity we define the transverse momentum conservation factor 
\begin{equation}
		\Pi_2(\{q_i\}) = (2\pi)^2 \delta^{(2)}(\pp-\kp_1 -\kp_2)(2\pi)^2 \delta^{(2)}(\pp-\bar{\kp}_1 -\bar{\kp}_2)
		\label{eq:momentumconservation}
\end{equation}
where $\{q_i\}$ denotes collectively the set of momenta $\{\pp,\kp_1,\kp_2,\bk_1,\bk_2\}$ appearing in the delta functions.

Since we are working in the dilute-dilute limit of the CGC, the information of the targets enter the gluon yields through two-point source correlators, 
$\langle\rho^\dagger_{A,a}(\kp_1)\rho_{A,a}(\bk_1)\rangle$ for $A$ and
$\langle\rho^\dagger_{B,b}(\kp_2)\rho_{B,b}(\bk_2)\rangle$ for $B$. The
amplitude momenta $\kp_1,\kp_2$ and the conjugate momenta $\bk_1,\bk_2$ are different, representing momentum transfers from the targets in the amplitude and conjugate amplitude. Each target insertion corresponds to a gluon field vertex, and for this reason each source appears dressed by a transverse gluon propagator. Furthermore,  $\Pi_2(\{q_i\})$
 enforces transverse momentum conservation independently in the amplitude and in its conjugate. 
The hard factor is given by 
\begin{equation}
	\label{eq:hard-factor}
\begin{split}
h(\boldsymbol{k}_1, \boldsymbol{k}_2, \bar{\boldsymbol{k}}_1, \bar{\boldsymbol{k}}_2)
&= \frac{4}{\pp^2}
\left[
\left(\delta^{ij}\delta^{\kappa\lambda}
+ \epsilon^{ij}\epsilon^{\kappa\lambda}\right)
k_1^i\, k_2^j\, \bar{k}_1^\kappa\, \bar{k}_2^\lambda
\right]
\end{split}
\end{equation}
which in the limit of $\kp_{1,2}=\bk_{1,2}$, one obtains the square of the Lipatov effective vertex~\cite{Lipatov:1995pn,Kovchegov:1997ke,Kovner:1995ja,Kovner:1995ts,Gunion:1981qs}.
The numerator in \cref{eq:hard-factor} contracts with the $2$D identity
$\epsilon^{ij}\epsilon^{\kappa\lambda}=\delta^{i\kappa}\delta^{j\lambda}-\delta^{i\lambda}\delta^{j\kappa}$, to give three scalar products,
\begin{equation}
	\label{eq:hard-factor-contracted}
	\begin{split}
		h(\boldsymbol{k}_1, \boldsymbol{k}_2, \bar{\boldsymbol{k}}_1, \bar{\boldsymbol{k}}_2)
		&= \frac{4}{\pp^2} \left[ (\kp_1 \cdot \kp_2) (\bk_1 \cdot \bk_2) + (\kp_1 \cdot \bk_1) (\kp_2 \cdot \bk_2) - (\kp_1 \cdot
		\bk_2) (\bk_1 \cdot \kp_2) \right]
	\end{split}
\end{equation}

Using the operator definition of unintegrated gluon distribution functions (uGDFs), one can get the    relation between the color source correlations and the space dependent  uGDF in the dilute-dilute limit~\cite{Blaizot:2004wu},  
\begin{equation}
	\begin{split}
		g^2\left\langle      \tilde{\rho}^\dagger_{A,a}(\kp_1)     \tilde{\rho}_{A,a}(\bk_1)
		\right\rangle_{x}\,
		=&\;\frac{1}{\pi}   \frac{\kp_1^2\bk_1^2} { \kp_1\cdot \bk_1} \int_{\xv} \rme^{-\rmi \xv \cdot (\kp_1-\bk_1) }\varphi \left(x_A, \frac{\kp_1+\bk_1}{2},\xv\right)\,.
	\end{split}
	\label{eq:wwfunction}
\end{equation}
where we have introduced, as it is standard, longitudinal dependence into the  gluon unintegrated functions through the longitudinal momentum fraction $x_{A,B}$ which in this channel can be obtained through kinematics as $x_{A,B}=p_\perp\,\rme^{\pm y}/\sqrt{s}$. 
Inserting \cref{eq:wwfunction} into \cref{eq:yieldsr} we get 

\begin{equation}
	\begin{split}
		\label{eq:yields_with_ww}
\frac{\rmd N_g}{\rmd^2 \pp\,\rmd y }
		&= \frac{g^2}{(2\pi)^5 C_F }
		\int_{\boldsymbol{k}_1 \bar{\boldsymbol{k}}_1 \boldsymbol{k}_2 \bar{\boldsymbol{k}}_2}\int_{\boldsymbol{x}\boldsymbol{y}}
		\;	\Pi_2(\{q_i\})\,
		\frac{h\left(\boldsymbol{k}_1,\boldsymbol{k}_2,\bar{\boldsymbol{k}}_1,\bar{\boldsymbol{k}}_2\right)}{(\kp_1\cdot\bk_1)(\kp_2\cdot\bk_2)} \\
		& \qquad  \qquad \times 		
		\mathrm{e}^{-\mathrm{i}\boldsymbol{x}\left(\boldsymbol{k}_1 - \bar{\boldsymbol{k}}_1\right)}
		\mathrm{e}^{-\mathrm{i}\boldsymbol{y}\left(\boldsymbol{k}_2 - \bar{\boldsymbol{k}}_2\right)} 
		\varphi_A\!\left(x_A, \tfrac{\boldsymbol{k}_1 + \bar{\boldsymbol{k}}_1}{2},\xv\right)
		\varphi_B\!\left(x_B, \tfrac{\boldsymbol{k}_2 + \bar{\boldsymbol{k}}_2}{2},\boldsymbol{y}\right)
	\end{split}
\end{equation}

For simplicity, we define the full prefactor as 
\begin{equation}
	\label{eq:overall_expansion_factor}
 \frac{h\left(\boldsymbol{k}_1,\boldsymbol{k}_2,\bar{\boldsymbol{k}}_1,\bar{\boldsymbol{k}}_2\right)}{(\kp_1\cdot\bk_1)(\kp_2\cdot\bk_2)}\equiv\frac{4}{\pp^2} H\left(\boldsymbol{k}_1,\boldsymbol{k}_2,\bar{\boldsymbol{k}}_1,\bar{\boldsymbol{k}}_2\right)
\end{equation}
where we have pulled the $4/\pp^2$ out so that the $H$ factor is dimensionless. Notice that here the combinations  $\kp_{1,2}- \bk_{1,2}$ 
are the Fourier conjugates of the spatial dependence within the uGDFs in \cref{eq:yields_with_ww}, and appear exclusively as variables within the hard factor. It is easier to work this out by performing a basis rotation to a Wigner momentum basis, 
\begin{equation}
\frac{\boldsymbol{k}_1 + \bar{\boldsymbol{k}}_1}{2} = \boldsymbol{q}_\perp,
\qquad
\frac{\boldsymbol{k}_2 + \bar{\boldsymbol{k}}_2}{2} = \boldsymbol{k}_\perp,
\qquad
\tilde{\boldsymbol{q}}_\perp \equiv \boldsymbol{k}_1 - \bar{\boldsymbol{k}}_1,
\qquad
\tilde{\boldsymbol{k}}_\perp \equiv \boldsymbol{k}_2 - \bar{\boldsymbol{k}}_2
	\label{eq:wigner_rotation}
\end{equation}
we can write the yield in a concise way, 
\begin{equation}
\begin{split}
\frac{\rmd N_g}{\rmd^2 \pp\,\rmd y }
	=&  \frac{\as}{2\pi^4 \pp^2 C_F}
	\int_{\boldsymbol{q}\tilde{\boldsymbol{q}}\boldsymbol{k}\tilde{\boldsymbol{k}}} \int_{\xv\yv}
	\Pi_2(\{q_i\})\,
	H\left(\boldsymbol{q}+\frac{1}{2}\tilde{\qp},\boldsymbol{k}+\frac{1}{2}\tilde{\kp},\boldsymbol{q}-\frac{1}{2}\tilde{\qp},\boldsymbol{k}-\frac{1}{2}\tilde{\kp}\right)\\
	& \qquad \quad\times \mathrm{e}^{-\mathrm{i}\tilde{\boldsymbol{q}}\cdot x}\,
	\mathrm{e}^{-\mathrm{i}\tilde{\boldsymbol{k}}\cdot y}\,
	\varphi_A(x_A,\boldsymbol{q},\xv)\,\varphi_B(x_B,\boldsymbol{k},\yv)
\end{split}
\label{eq:pretty}
\end{equation}

The hard factor $H$ in \cref{eq:pretty}, together with the four eikonal denominators, is a smooth function of the shifted momenta $\qp\pm\qtil/2,\kp\pm\ktil/2$.  Since $\qtil,\ktil$ are the Fourier-conjugate variables to the average position of the sources, they are parametrically of order the inverse of the length scale $R$ over which the color-charge densities vary. On the other hand, 
 the mean momenta $\qp,\kp$ probe the characteristic momentum scale of the distribution, and hence are larger than the saturation scale $Q_s$ in the dilute-dilute limit.
  For distributions which vary in transverse position slowly enough that $Q_s R\gg1$, each power of $\qtil/\qp$ or $\ktil/\kp$ carries a further suppression factor $ 1/(Q_s R)\ll 1$. We can then  reorganize $H$ by expanding order by order in the relative momenta. Since the previously mentioned computations worked in the $R\to\infty$ limit, we can identify their results to the zeroth order of such expansion. In what follows we will also explicitly show this.
By keeping the full tower of terms, and collecting all contributions of order $n$ in $\qtil$ and $m$ in $\ktil$ defines a set of tensor coefficients
\begin{equation}
	H\left(\boldsymbol{k}_1,\boldsymbol{k}_2,\bar{\boldsymbol{k}}_1,\bar{\boldsymbol{k}}_2\right)=\sum_{n,m=0}^{\infty}\frac{1}{n!\,m!}\tilde{q}^{i_1}...\tilde{q}^{i_n} \tilde{k}^{j_1} ... \tilde{k}^{j_m}\,H_{(n,m)}^{i_1,...,i_n;j_1,...,j_m}(\qp,\kp).
\end{equation}
here $H_{(n,m)}^{i_1\dots i_n;j_1\dots j_m}(\qp,\kp)$ represents the $n$-th and $m$-th transverse derivatives of $H$ at $\qtil=\ktil=0$. It is useful to reorganize the expansion term by grouping all contributions to a certain order $l$, $H=\sum_{l=0}^{\infty}H_l$, where
\begin{equation}
	H_l(\qp,\kp;\qtil,\ktil)=\sum_{n,m=0}^{\infty}\frac{\delta_{l,n+m}}{n!\,m!}\tilde{q}^{i_1}...\tilde{q}^{i_n} \tilde{k}^{j_1} ... \tilde{k}^{j_m}\,H_{(n,m)}^{i_1,...,i_n;j_1,...,j_m}(\qp,\kp).
\end{equation}
 Inserting this expression into the particle production formula, we get 
\begin{equation}
	\frac{\rmd N_g}{\rmd^2 \pp\,\rmd y }
	= \sum_{l=0}^{\infty}\frac{\as}{2\pi^4 \pp^2 C_F}
	\int_{\qp\qtil\kp\ktil}\int_{\xv\yv}	\Pi_2(\{q_i\})\,
	H_l(\qp,\kp,\qtil,\ktil) \,
	\rme^{-\rmi\qtil\cdot \xv}\rme^{-\rmi\ktil\cdot \yv}\,
	\varphi_A(x_A,\qp,\xv)\varphi_B(x_B,\kp,\yv)
\end{equation}

Since every term $H_l(\qp,\kp,\qtil,\ktil)$ corresponds to a polynomial term in $\qtil$ and $\ktil$, we can use the plane wave terms to exchange the relative momenta for transverse space derivatives, $\qtil \to -\rmi \nabla_{\xv}$ and $\ktil \to -\rmi \nabla_{\yv}$, where, naturally, $\nabla_i$ corresponds to the transverse-space gradient with respect to variable $i$. Integrating by parts, and assuming that the color charge density vanishes far away from the center of the targets, $\varphi_{A}|_{|\xv|\to \infty} \to 0$ and $\varphi_{B}|_{|\yv|\to \infty} \to 0$, we can express our formula as a proper gradient expansion. The ladder of contributions is given by
\begin{equation}
	\begin{split}
		\frac{\rmd N_g}{\rmd^2 \pp\,\rmd y }
		=& \sum_{l=0}^{\infty}\frac{\as}{2\pi^4 \pp^2 C_F}
		\int_{\qp\qtil\kp\ktil}\int_{\xv\yv}	\Pi_2(\{q_i\})\,\rme^{\,\rmi\qtil\cdot \xv}\rme^{\,\rmi\ktil\cdot \yv}\\
		&\quad\qquad\times \left[H_l(\qp,\kp,-\rmi \nabla_{\xv},-\rmi \nabla_{\yv})\,
		\varphi_A(x_A,\qp,\xv)\varphi_B(x_B,\kp,\yv)\right]
	\end{split}
\end{equation}

We can get rid of the momentum conservation factor, $\Pi_2(\{q_i\})$, by using the fact that the term in square brackets does not depend anymore in the releative momenta. By integrating $\ktil$ using the delta functions we obtain $\ktil = 2(\pp-\qp-\kp) - \qtil$, which after reinserting in the expression for $\Pi_2(\{q_i\})$, renders
\begin{equation}
	\begin{split}
			\int_{\qp\qtil\kp\ktil} \rme^{\,\rmi\qtil\cdot \xv}\rme^{\,\rmi\ktil\cdot \yv}\,	\Pi_2(\{q_i\})\, &= 
		\int_{\qtil} \rme^{\,\rmi\qtil\cdot (\xv-\yv)}\,
		(2\pi)^2 \delta^{(2)}\left(\pp-\qp-\kp\right)\\
		& = (2\pi)^2 \delta^{(2)}\left(\pp-\qp-\kp\right) \delta^{(2)}\left(\xv-\yv\right)
	\end{split}
\end{equation} 
We get the final expression for gluon production 
\begin{equation}
	\begin{split}
		\frac{\rmd N_g}{\rmd^2 \pp\,\rmd^2 \xv\,\rmd y }
		=& \sum_{l=0}^{\infty}\frac{\as}{2\pi^4 \pp^2 C_F}
		\int_{\qp\kp}(2\pi)^2 \delta^{(2)}\left(\pp-\qp-\kp\right)\\
		&\qquad \times \left[H_l(\qp,\kp,-\rmi \nabla_{\xv}, -\rmi \nabla_{\yv})\,
		\varphi_A(x_A,\qp,\xv)\varphi_B(x_B,\kp,\yv)\right]_{\yv=\xv}\\
		=&\sum_{l=0}^{\infty} E_p\frac{\rmd N^{(l)}_g}{\rmd^3 p\, \rmd^2 \xv}
	\end{split}
	\label{eq:order-by-order}
\end{equation}
 
In \cref{eq:order-by-order}, $H_l$ acts as an $l$-th order differential operator on $\varphi_A,\varphi_B$ before setting $\yv=\xv$. Additionally, notice that while yields are by construction positive in total, each $l$th term does not need to be positive definite. While the last line of \cref{eq:order-by-order} is not technically a sum of distributions, it is really useful notation to keep to denote the corrections to the leading/standard $k_\perp$-factorized formula. In this section we will work up to the second non-trivial order correction. 

\paragraph{Zeroth order:} The lowest level contains no relative momenta in the numerator, $H_0= 1$, and therefore 
\begin{equation}
	\begin{split}
		E_p\frac{\rmd N^{(0)}_g}{\rmd^3 p\, \rmd^2 \xv}= \frac{\as}{2\pi^4 \pp^2 C_F}
		\int_{\qp\kp}(2\pi)^2 \delta^{(2)}\left(\pp-\qp-\kp\right)
		\varphi_A(x_A,\qp,\xv)\varphi_B(x_B,\kp,\xv)\,,
		\label{eq:ktfactorization}
	\end{split}
\end{equation}
 which recovers the standard \ktfact{} formula. 
At this order the two targets are treated as locally homogeneous at $\xv$, exactly as if they were infinite and translationally invariant.
Physically, the emitted gluons cannot resolve the spatial variation of the uGDFs in this limit, as their spatial size $1/\pT$ is too small in comparison with the radius of change in the color densities. All of the dependence on the finite size of the sources is therefore contained in the higher orders, to which we turn next.

\paragraph{Leading order correction:} As stated before, the next non-trivial order is of second order in relative momenta (or gradients). The numerator is given by a sum of  tensor structures contracted with the relative momenta, 
\begin{equation}
	H_2(\qp,\kp;\qtil,\ktil)=-\frac{1}{\kp^2\qp^2} \tilde{k}^i   P^{ij}_{\qp\kp}\tilde{q}^j\quad \text{with} \quad  P^{ij}_{\qp\kp} = q^i k^j -\delta^{ij} \qp\cdot\kp
	 \label{eq:H2}
\end{equation}

Notice that the first order seems to diverge for values $|\qp|,|\kp|\to 0$. Together with the respective jacobians in the $\qp, \kp$ integration, the numerator cancels the denominator, and the whole correction formula is finite at small values of the incoming gluon momenta, within the correction factor.  Substituting the formula for gradients, we get a compact formula for the first gradient correction to the $k_\perp$-factorization formula

\begin{equation}
	\begin{split}
		\frac{\rmd N_g}{\rmd^2 \pp\,\rmd^2 \xv\,\rmd y }
		=& \frac{\as}{2\pi^4 \pp^2 C_F}
		\int_{\qp\kp}(2\pi)^2 \delta^{(2)}\left(\pp-\qp-\kp\right)
	\frac{P^{ij}_{\qp\kp}}{\kp^2\qp^2}  	\left[
		 \nabla_{\xv}^j \varphi_A(x_A,\qp,\xv)\nabla_{\yv}^i  \varphi_B(x_B,\kp,\yv)\right]_{\yv=\xv}
	\end{split}
	\label{eq:first_order}
\end{equation}
To understand the consequences of this first non-trivial term is useful to recast the equation simply as the spatial gradients of a tensor structure, 
\begin{equation}
		\frac{\rmd N_g}{\rmd^2 \pp\,\rmd^2 \xv\,\rmd y }
		= \Big[
		\nabla_{\yv}^i\nabla_{\xv}^j\big(I_2^{ij}-\delta^{ij}\tr I_2\big)
		\Big]_{\yv=\xv},
	\label{eq:LOpretty}
\end{equation} 
The integral under the gradients can be simply expressed as 
\begin{align}
	I_{2}^{ij}(\pp,\xv,\yv) &=\frac{\as}{2\pi^4 \pp^2 C_F} \int_{\qp\kp}(2\pi)^2\delta(\pp-\qp-\kp)\;
	q^i k^j\;\frac{\varphi_A(x_A,\xv,\qp)}{\qp^2}\frac{\varphi_B(x_B,\yv,\kp)}{\kp_\perp^2}.
\end{align}
This  integral construction makes manifest how anisotropies appear in the gradient expansion, as due to symmetries we can express
\begin{equation}
	I_{2}^{ij}(\pp,\xv,\yv) = A(\pp,\xv,\yv) p^{i} p^{j} + B(\pp,\xv,\yv) \delta^{ij}
	\label{eq:decomposition_I1}
\end{equation}

Even if we chose $\varphi_{A,B}$ to be isotropic, finite sources, the coupling between the outgoing gluon momentum and the gradient would not vanish, as the radial direction gives a preferential direction, singled out by the density gradients at $\xv,\yv$, so the correction depends on the angle between $\pp$ and that direction. The coupling becomes more intricate when taking into account 
a non-vanishing impact parameter between the incoming nuclei. 
In this case, even the radially symmetric gluon distributions would create a non-trivial interacting geometry, and allowing for non-cancelations between the gradients of the respective uGDFs\footnote{For readers interested in explicit realizations of such geometries, please refer to the accompanying phenomenological paper \cite{Garcia-Montero:2026NextPaper}}.  
The anisotropies produced in \cref{eq:LOpretty} are then a property  generated dynamically by this coupling of the two sources with the outgoing gluon. Inserting \cref{eq:decomposition_I1} into \cref{eq:LOpretty}, we get
\begin{equation}
	\frac{\rmd N_g}{\rmd^2 \pp\,\rmd^2 \xv\,\rmd y }
	= \left\{
	\nabla_{\yv}^i\nabla_{\xv}^j\left[A(\pp,\xv,\yv) \left( p^{i} p^{j}-\delta^{ij}\pp^2\right) -  \delta^{ij}\,B(\pp,\xv,\yv)
	\right]
	\right\}_{\yv=\xv},
	\label{eq:LOpretty_aniso_explicit}
\end{equation}
Written this way, the correction splits into an anisotropic and an isotropic term. The second term, generated by
$B$ is diagonal, and hence contracts  $\nabla_{\yv}^i\nabla_{\xv}^i \varphi_A \varphi_B$ into a single scalar with no
reference to the direction of the produced gluon. Therefore this term will not be able to carry any anisotropy. The effect is a rescaling of the local yield relative to the zeroth-order value
\cref{eq:ktfactorization}. The value of this rescaling is by how steeply and 
coherently the two overlapping profiles vary at a local point $\xv$. In contrast, the $A$ term contains $P^{ij}_{\pp\pp}\nabla^{i}_{\yv}\nabla^{j}_{\xv}$. Using the 2D Levi-Civita identity, we can transform this term into $(\pp\times \nabla_{\xv})(\pp\times \nabla_{\yv})$, where each factor projects onto the direction orthogonal to the momentum of the emitted gluon.

As we will see, this has a direct consequence for the energy-momentum tensor of the produced gluons, \cref{eq:order_by_order_Tmunu}. While the isotropic part of the number density, $B\,\delta^{ij}$ integrates to simplyshift the local energy (and hence the isotropic \ktfact{} result), the $A\,p^ip^j$ term feeds into higher moments of the pressure tensor. Wherever $\nabla_{\xv}\varphi_A$ and $\nabla_{\yv}\varphi_B$ are locally non-zero, this generates a transverse pressure imbalance $T^{xx}-T^{yy}\neq 0$ together with a shear component $T^{xy}$, sourced entirely by the spatial inhomogeneity of the two targets at the production point of the gluon.

\paragraph{Next to leading order:} 
The full quartic numerator $H_4$, is found to be surprisingly simple, namely
\begin{equation}
	\begin{split}
			H_4(\qp,\kp,\qtil,\ktil)  =&  -\frac{\tilde{k}^i   P^{ij}_{\qp\kp}\tilde{q}^j}{4(\qp^2\kp^2)^2} \left[\ktil^2 \qp^2 + \qtil^2 \kp^ 2\right]
	\end{split}
	\label{eq:H4}
\end{equation}
Note that $H_4$ carries exactly the same anisotropic structure 
found at first order,  $P^{ij}_{\qp\kp}$. The interpretation from the second order correction holds for these terms, with the caveat that we have now $A\leftrightarrow B$ symmetry breaking terms, represented by the inclusion of extra Laplacian operators. This is why, at order $l=4$,the emitted gluons correlate to the joint gradients of $\varphi_A\nabla^2_{\yv} \varphi_B$ (on the left term), which has a richer structure, but is suppressed by two powers of $R^{-2}$. The existence of the two Laplacian terms in $H_4$ restores $\qp,\qtil \leftrightarrow\kp\ktil$ and hence the  $A,B$ symmetry for equal ion collisions. However what is interesting is that within the quadratic factors, each term induces an odd number of powers of $\tilde{q}$ and of $\tilde{k}$ individually (one and three, respectively, in the first term), leading to directional asymmetries once the color charge is evaluated event-by-event (EbE). 
Substituting $\qtil\to\rmi\nabla_{\xv}$, $\ktil\to\rmi\nabla_{\yv}$ as before gives the corresponding gradient-expansion term,
\begin{equation}
	\begin{split}
		\frac{\rmd N_g}{\rmd^2 \pp\,\rmd^2 \xv\,\rmd y }
		=& -\frac{\as}{2\pi^4\pp^2 C_F}
		\int_{\qp\kp}(2\pi)^2\delta^{(2)}\left(\pp-\qp-\kp\right)
		\frac{P^{ij}_{\qp\kp}}{4(\qp^2\kp^2)^2}\\
		&\times\Big[\qp^2\,\nabla_{\xv}^j\varphi_A(x_A,\qp,\xv)\,\nabla_{\yv}^i\nabla_{\yv}^2\varphi_B(x_B,\kp,\yv)\\
		&\qquad+\kp^2\,\nabla_{\xv}^j\nabla_{\xv}^2\varphi_A(x_A,\qp,\xv)\,\nabla_{\yv}^i\varphi_B(x_B,\kp,\yv)\Big]_{\yv=\xv}
	\end{split}
	\label{eq:second_order}
\end{equation}
where the transverse laplacian $\nabla_{\xv}^2\equiv \delta^{ij}\nabla_{\xv}^i\nabla_{\xv}^j$. As in the case of the first non-vanishing order, each term pairs one gradient of $\varphi_A$ with a Laplacian-and-gradient of $\varphi_B$, or vice versa, so the correction still vanishes for a pair of exactly homogeneous sources and is controlled by the same $1/(Q_sR)$ counting, now at one higher order in transverse gradients.

As before, we recast \cref{eq:second_order} as gradients acting on momentum integrals evaluated at $\yv=\xv$,
\begin{equation}
		\frac{\rmd N_g}{\rmd^2 \pp\,\rmd^2 \xv\,\rmd y }
	= -
	\Big[\nabla_{\yv}^i\nabla_{\yv}^2\nabla_{\xv}^j\big(I_4^{ij}-\delta^{ij}\tr I_4\big)
	+\nabla_{\yv}^i\nabla_{\xv}^j\nabla_{\xv}^2\big(\bar{I}_4^{ij}-\delta^{ij}\tr \bar{I}_4\big)\Big]_{\yv=\xv},
	\label{eq:NLOpretty}
\end{equation}
with
\begin{align}
	I_4^{ij}(\pp,\xv,\yv) &=\frac{\as}{8\pi^4\pp^2 C_F} \int_{\qp\kp}(2\pi)^2\delta(\pp-\qp-\kp)\;
	\frac{q^ik^j}{\kp^2}\,\frac{\varphi_A(x_A,\xv,\qp)}{\qp_\perp^2}\frac{\varphi_B(x_B,\yv,\kp)}{\kp_\perp^2},\\
	\bar{I}_4^{ij}(\pp,\xv,\yv) &=\frac{\as}{8\pi^4\pp^2 C_F} \int_{\qp\kp}(2\pi)^2\delta(\pp-\qp-\kp)\;
	\frac{q^ik^j}{\qp^2}\,\frac{\varphi_A(x_A,\xv,\qp)}{\qp_\perp^2}\frac{\varphi_B(x_B,\yv,\kp)}{\kp_\perp^2}.
\end{align}
It is worth noting that $I_4^{ij}$ and $\bar{I}_4^{ij}$ are the same integral, differing only in which of the two eikonal denominators carries the extra power; exchanging $\qp\leftrightarrow\kp$ (equivalently $A\leftrightarrow B$) in ${I_4^{ij}}_{A\leftrightarrow B}=\bar{I}_4^{ji}$. The reader can then notice that the momentum integrals $I_2$ and $I_{4}$  are higher moments of
the same unintegrated distributions \cref{eq:ktfactorization}, weighted by powers
of sampled gluons' momenta. Therefore, successive gradient corrections therefore will probe finer features of the target's transverse phase-space density.

Finally, we can get the energy-momentum tensor using \cref{eq:Tmunu}, and the gradient expansion \cref{eq:order-by-order}. For this, it is useful to define the $n$th gradient correction to the energy-momentum tensor via\begin{equation}
	\tau T^{\mu\nu} _{n}= \int \rmd ^2 \pp \,    \frac{p^{\mu}p^{\nu} }{ p^\tau}\,E_p\frac{\rmd N^{(n)}}{\rmd^3 p\,\rmd^2 \xv}\bigg|_{y=\eta }
	\label{eq:order_by_order_Tmunu}
\end{equation}
In this case we can get an analogous formula to the one first and second order terms for the gluon yields
\begin{equation}
	\begin{split}
		\tau T^{\mu\nu} _{2} &=\Bigg[
		\nabla_{\yv}^i\nabla_{\xv}^j\bigg(\calH_2^{\mu\nu,ij}-\delta^{ij} \calH_2^{\mu\nu,kl} \delta_{kl}\bigg)
		\Bigg]_{\yv=\xv},\\
		\tau T^{\mu\nu} _{4} &=-
		\Bigg[\nabla_{\yv}^i\nabla_{\yv}^2\nabla_{\xv}^j\bigg(\calH_4^{\mu\nu,ij}-\delta^{ij} \calH^{\mu\nu,kl}_4\delta_{kl}\bigg)
		+\nabla_{\yv}^i\nabla_{\xv}^j\nabla_{\xv}^2\bigg(\bar\calH_4^{\mu\nu,ij}-\delta^{ij}\bar\calH^{\mu\nu,kl}_4 \delta_{kl}\bigg)\Bigg]_{\yv=\xv}
		\label{eq:Tmunu_os_final}
	\end{split}
\end{equation} 
where the functions needed to compute $\tau T^{\mu\nu}_n$ are given by 
\begin{equation}
	\begin{split}
		\calH_n^{\mu\nu,ij}(\xv,\yv)&=  \int \rmd ^2 \pp \,    \frac{p^{\mu}p^{\nu} }{ p^\tau}\, I^{ij}_n(\pp,\xv,\yv) \\
	\end{split}
\end{equation}

These formulas are particularly useful for a phenomenological realization of
the initial conditions in heavy-ion collisions. The gradient corrections
are local operators built from transverse derivatives of these integrals, which are nothing but complex moments of the unintegrated
distributions. This means that these terms can be added order by order on top of any saturation-based
initial-state model that already implements the zeroth-order
$k_\perp$-factorized formula, such as
the \Dipper{} model~\cite{Garcia-Montero:2023gex,garcia_montero_2026_20179552}, at the cost
of a few extra derivatives of the source profiles per order. Alternatively, a phenomenologically ready parametric formula can be obtained using the techniques of Refs.~\cite{Borghini:2022iym,Garcia-Montero:2024jev}, where these new terms act as anisotropy-corrected versions of a CGC TRENTo-like energy density.
 Such a formula and its
phenomenological consequences for light-ion collisions, will be developed from the GBW model in the companion paper~\cite{Garcia-Montero:2026NextPaper}.

\section{The dilute-dilute limit of the CGC and $k_\perp$-factorization: real-time evolution}
\label{sec:real_time}
The gradient expansion of \cref{sec:tmunu} was carried out on the on-shell gluon
spectrum, converted to $T^{\mu\nu}$ via a kinetic picture, which treats
the produced gluons as free on-shell quanta with $y=\eta$. By construction, this method 
misses some infrared physics due to dynamics of classical color fields. For example, the coherent interference between
the amplitude and its conjugate at different mode momenta, which yields terms including chromo-electric and -magnetic interference. Since the early-time matter does behave as a classical field,  the dilute Glasma in our setting, we will repeat the dilute--dilute computation in
position space from the results for a  classical field strength of the Vienna group, see
Refs.~\cite{Ipp:2021lwz,Ipp:2024ykh}. We recover the same gradient expansion,
with the momentum-space result of \cref{sec:tmunu} as the leading orders
of a complex ladder of terms contributing to the energy-momentum tensor of the early stages of a heavy-ion collision. 

The starting point is the covariant-gauge color field of each single nucleus and
the perturbation it sources when the two overlap. The incoming CGC fields,
are given by \cref{eq:BG_fields} and \cref{eq:BG_fields_FT} in transverse Fourier space. Generally, the full solution for the classical gauge fields can be obtained by inverting the CYM equations, where the interacting gauge field produced by the collision can be obtained by separating it from the original background fields $A_{A,B}^{\mu}$, via $
	A^{\mu} = A_A^{\mu}  + A_B^{\mu} + a^{\mu} $
where $a^\mu$ is the interaction piece. In the dilute-dilute limit we restrict the inversion of the CYM equation so that $a^\mu$ only contains first order insertions of the sources, i.e $a^\mu \sim O[\rho_A \rho_B]$. The perturbation field is given then by \cite{Ipp:2021lwz,Ipp:2024ykh} 
\begin{equation}
	\begin{split}
			a^\mu(x) =& \;\frac{g}{2}\, f_{abc}\, t^c
		\int_{\kp_1,\kp_2} \int_0^\infty \rmd v^+ \int_0^\infty \rmd v^-\; \tilde{a}^{\mu}(v; \kp_1,\kp_2) 	\rme^{-\rmi(\boldsymbol{k}_1+\boldsymbol{k}_2)\cdot\xv}\,, \\
	& \times	\tilde{\mathcal{A}}_{A,a}^{-}\!\left(x^+ - v^+, \boldsymbol{k}_1\right)\,
		\tilde{\mathcal{A}}_{B,b}^{+}\!\left(x^- - v^-, \boldsymbol{k}_2\right)\\
		\label{eq:a_mu}
	\end{split}
\end{equation}
where the components of the integration kernel are given by  
\begin{equation}
	\begin{split}
		\tilde{a}^\pm(x) =&  \frac{\left( \kp_2^2 -\kp_1^2 \mp 2 \kp_1 \cdot \kp_2 \right)}{|\boldsymbol{k}_1+\boldsymbol{k}_2|\,\tau'}\,
		\,v^\pm J_1\!\left(|\boldsymbol{k}_1+\boldsymbol{k}_2|\,\tau'\right)\\
		\tilde{a}^i(x) =& \left(k_1^i - k_2^i\right)
		J_0\!\left(|\boldsymbol{k}_1+\boldsymbol{k}_2|\,\tau'\right)
	\end{split}
\end{equation}
from which one can get the field strength tensor  by using ${\mathcal{F}}^{\mu\nu} = \partial^\mu a^\nu- \partial^\nu a^\mu$. Component by component, one gets 
\begin{align}
	\label{eq:fieldstrength_raw}
	\mathcal{F}^{+-} =& \;g\, f_{abc}\, t^c
	\int_{\kp_1,\kp_2} \int_0^\infty \rmd v^+ \int_0^\infty \rmd v^-\;
	\tilde{\mathcal{A}}_{A,a}^{-}\!\left(x^+ - v^+, \boldsymbol{k}_1\right)\,
	\tilde{\mathcal{A}}_{B,b}^{+}\!\left(x^- - v^-, \boldsymbol{k}_2\right)\notag\\
	& \times (\boldsymbol{k}_1\cdot\boldsymbol{k}_2)\,
	J_0\!\left(|\boldsymbol{k}_1+\boldsymbol{k}_2|\,\tau'\right)\,
	\rme^{-\rmi(\boldsymbol{k}_1+\boldsymbol{k}_2)\cdot\xv}\,,\\
	\mathcal{F}^{+i} =& -\rmi g\, f_{abc}\, t^c
	\int_{\kp_1,\kp_2} \int_0^\infty \rmd v^+ \int_0^\infty \rmd v^-\;
	\tilde{\mathcal{A}}_{A,a}^{-}\!\left(x^+ - v^+, \boldsymbol{k}_1\right)\,
	\tilde{\mathcal{A}}_{B,b}^{+}\!\left(x^- - v^-, \boldsymbol{k}_2\right)\notag\\
	& \times \frac{v^+}{|\boldsymbol{k}_1+\boldsymbol{k}_2|\,\tau'}
	\left[\,k_1^i\,\boldsymbol{k}_2^2 - k_2^i\left(\boldsymbol{k}_1^2 + 2(\boldsymbol{k}_1\cdot\boldsymbol{k}_2)\right)\right]
	J_1\!\left(|\boldsymbol{k}_1+\boldsymbol{k}_2|\,\tau'\right)\,
	\rme^{-\rmi(\boldsymbol{k}_1+\boldsymbol{k}_2)\cdot\xv}\,,\\
	\mathcal{F}^{-i} =& -\rmi g\, f_{abc}\, t^c
	\int_{\kp_1,\kp_2} \int_0^\infty \rmd v^+ \int_0^\infty \rmd v^-\;
	\tilde{\mathcal{A}}_{A,a}^{-}\!\left(x^+ - v^+, \boldsymbol{k}_1\right)\,
	\tilde{\mathcal{A}}_{B,b}^{+}\!\left(x^- - v^-, \boldsymbol{k}_2\right)\notag\\
	& \times \frac{v^-}{|\boldsymbol{k}_1+\boldsymbol{k}_2|\,\tau'}
	\left[\,k_1^i\left(\boldsymbol{k}_2^2 + 2(\boldsymbol{k}_1\cdot\boldsymbol{k}_2)\right) - k_2^i\,\boldsymbol{k}_1^2\right]
	J_1\!\left(|\boldsymbol{k}_1+\boldsymbol{k}_2|\,\tau'\right)\,
	\rme^{-\rmi(\boldsymbol{k}_1+\boldsymbol{k}_2)\cdot\xv}\,,\\
	\mathcal{F}^{ij} =& -g\, f_{abc}\, t^c
	\int_{\kp_1,\kp_2} \int_0^\infty \rmd v^+ \int_0^\infty \rmd v^-\;
	\tilde{\mathcal{A}}_{A,a}^{-}\!\left(x^+ - v^+, \boldsymbol{k}_1\right)\,
	\tilde{\mathcal{A}}_{B,b}^{+}\!\left(x^- - v^-, \boldsymbol{k}_2\right)\notag\\
	& \times \left(k_2^i k_1^j - k_2^j k_1^i\right)
	J_0\!\left(|\boldsymbol{k}_1+\boldsymbol{k}_2|\,\tau'\right)\,
	\rme^{-\rmi(\boldsymbol{k}_1+\boldsymbol{k}_2)\cdot\xv}\,.
\end{align}

We will cast these expressions towards the large time limit. This can be done using the Bessel function's large argument asymptotic limit 
\begin{equation}
	\label{eq:asymptoticBessel}
	J_\alpha(z) \sim \sqrt{\frac{2}{\pi z}}\,
	\cos\left(z - \frac{\alpha\pi}{2} - \frac{\pi}{4}\right)
	\implies
	\begin{cases}
		J_0(z) \sim \dfrac{1}{\sqrt{2\pi z}}\left(\zeta_0\,\rme^{\rmi z } + \zeta_0^{*}\,\rme^{-\rmi z }\right), \\[8pt]
		J_1(z) \sim \dfrac{1}{\sqrt{2\pi z}}\left(-\rmi \zeta_0\,\rme^{\rmi z } + \rmi \zeta_0^{*}\,\rme^{-\rmi z }\right).
	\end{cases}
\end{equation}
where  $\zeta_0 = \exp(-\rmi \pi/4)$, which is valid when $|\kp_1+\kp_2|\tau \gg 1$. While this may naively look like a large time approximation, this is in fact a large momentum mode approximation. Furthermore, as we are working in the dilute limit, and the fields are expected to present $p_\perp \gtrsim Q_s$, this expansion is fully consistent with the initial truncation in the sources. Inserting in \cref{eq:fieldstrength_raw} we get that the field tensor now keeps its reality, but separates into complex pieces, 
\begin{equation}
	\mathcal{F}^{\mu\nu}(x)= f^{\mu\nu}(x) + f^{\mu\nu,*}(x)
\end{equation} where we will use the following sleeker form for positive energy field strength tensor, namely
\begin{align}
	\label{eq:fieldstrength_large}
	f^{\mu\nu} =& \;\frac{g}{\sqrt{2\pi}}\, f_{abc}\, t^c \zeta_0	\int_{\kp_1,\kp_2} \int_0^\infty \rmd v^+ \int_0^\infty \rmd v^-\;
	\tilde{\mathcal{A}}_{A,a}^{-}\!\left(x^+ - v^+, \boldsymbol{k}_1\right)\,
	\tilde{\mathcal{A}}_{B,b}^{+}\!\left(x^- - v^-, \boldsymbol{k}_2\right)\notag\\
	& \times \tilde{f}^{\mu\nu}(v, \kp_1,\kp_2)\,
	\rme^{\rmi|\boldsymbol{k}_1+\boldsymbol{k}_2|\,\tau'}\,
	\rme^{-\rmi(\boldsymbol{k}_1+\boldsymbol{k}_2)\cdot\xv}\,.
\end{align}

The components $ \tilde{f}^{\mu\nu}(v, \kp_1,\kp_2)$ keep part of their complexity, and are explicitly given by
\begin{equation}
	\begin{split}
		\tilde{f}^{+-}(v, \kp_1,\kp_2) &= \frac{(\boldsymbol{k}_1\cdot\boldsymbol{k}_2)}{\sqrt{|\boldsymbol{k}_1+\boldsymbol{k}_2|\,\tau'}}\,,  \qquad  \tilde{f}^{ij}(v, \kp_1,\kp_2) = \frac{1}{\sqrt{|\boldsymbol{k}_1+\boldsymbol{k}_2|\,\tau'}}\,  \left(k_1^i k_2^j - k_1^j k_2^i\right)\\
		\tilde{f}^{+i}(v, \kp_1,\kp_2) &= -\frac{ v^+}{\left(|\boldsymbol{k}_1+\boldsymbol{k}_2|\,\tau'\right)^{3/2}}
		\left[\,k_1^i\,\boldsymbol{k}_2^2 - k_2^i\left(\boldsymbol{k}_1^2 + 2(\boldsymbol{k}_1\cdot\boldsymbol{k}_2)\right)\right]\\
		\text{and} \quad  
		\tilde{f}^{-i}(v, \kp_1,\kp_2) &=  -\frac{ v^-}{\left(|\boldsymbol{k}_1+\boldsymbol{k}_2|\,\tau'\right)^{3/2}}
		\left[\,k_1^i\left(\boldsymbol{k}_2^2 + 2(\boldsymbol{k}_1\cdot\boldsymbol{k}_2)\right) - k_2^i\,\boldsymbol{k}_1^2\right]
	\end{split}
\end{equation}
The energy-momentum tensor is given by
\begin{equation}
	T^{\mu\nu} = 2\,\mathrm{Tr}\!\left[\,
	f^{\mu\rho}\,f_{\rho}{}^{\nu}
	+ \frac{1}{4}\,g^{\mu\nu}\,f^{\rho\sigma}\,f_{\rho\sigma}
	\,\right]
	\label{eq:Tmunu_space}
\end{equation}
Before we continue, let us take a look at two important elements in the construction of the energy-momentum tensor, and  compare the $(F^{\mu\rho})^* F_{\rho}^\nu $, and $F^{\mu\rho} F_{\rho}^\nu  $. From these pieces all of the construction follows, as the remaining pieces are either contractions or complex conjugated versions of them. The important element to realize is their dependence on the time-like phase. We can sketch their dependence as 
\begin{equation}
	(F^{\mu\rho})^* F_{\rho}^\nu  \sim 
	\rme^{-\rmi|\kp_1+\kp_2|\,\tau'}\,
	\rme^{\rmi|\bk_1+\bk_2|\,\bar{\tau}'}\,
	\quad\text{while}\quad 
	F^{\mu\rho} F_{\rho}^\nu  \sim \rme^{\rmi|\kp_1+\kp_2|\,\tau'}\,
	\rme^{\rmi|\bk_1+\bk_2|\,\bar{\tau}'}
\end{equation}

Since $\tau'$, $\bar{\tau}'$, and the norm of the vectors all are positive, and their combinations, the phase of the second term will always be large. Hence, by the Riemann-Lebesgue theorem,  in the limit in which we are working, the interference terms are vanishingly small. After a bit of algebra, the two surviving pieces give the expression for the energy-momentum tensor, 
\begin{equation}
	\label{eq:tmunu_fmunu}
	\begin{split}
		\langle T^{\mu\nu}(x)\rangle=&2\as\, f_{abc}f_{\bara \barb c}\, \int_{\kp_1,\kp_2}\;\int_{\bk_1,\bk_2}  \int_0^\infty \rmd v^+ \int_0^\infty \rmd v^- \int_0^\infty \rmd \bvp\int_0^\infty \rmd \bvm
		\\
		& \times \left\langle 	\tilde{\mathcal{A}}_{A,a}^{-,\dagger}\!\left(x^+ - v^+, \boldsymbol{k}_1\right)\,	\tilde{\mathcal{A}}_{A,\bara}^{-}\!\left(x^+ - \bvp, \bk_1\right)\, \right\rangle\\ 
		&\times \left\langle 	\tilde{\mathcal{A}}_{B,b}^{+,\dagger}\!\left(x^- - v^-, \boldsymbol{k}_2\right)\,	\tilde{\mathcal{A}}_{B,\barb}^{+}\!\left(x^- - \bvm, \bk_2\right)\, \right\rangle
	\end{split}
\end{equation}
We can define $t^{\mu\nu}$, an  symmetric energy-momentum kernel, by 
\begin{equation}
\begin{split}
		t^{\mu\nu}(v, \kp_1,\kp_2;\bar{v}, \bk_1,\bk_2)\, =& 
	\tilde{f}^{\mu\rho}(v, \kp_1,\kp_2)\,\tilde{f}_{\rho}{}^{\nu}(\bar{v}, \bk_1,\bk_2)+\tilde{f}^{\nu\rho}(v, \kp_1,\kp_2)\,\tilde{f}_{\rho}{}^{\mu}(\bar{v}, \bk_1,\bk_2)\\
	&+ \frac{1}{2}\,g^{\mu\nu}\,\tilde{f}^{\rho\sigma}(v, \kp_1,\kp_2)\,\tilde{f}_{\rho\sigma}(\bar{v}, \bk_1,\bk_2)
\end{split}
	\label{eq:tmunu}
\end{equation}
Notice that now, the integration kernel is symmetric under the exchange $(v,\kp_1,\kp_2)\leftrightarrow(\bar v, \bk_1,\bk_2)$, and hence real.
Assuming gaussianity in the color indices of the sources, and the fields, we can project their correlations in the following way, 
\begin{equation}
	\label{eq:phi_correlations}
	\begin{split}
		&\left\langle 	\tilde{\mathcal{A}}_{A,a}^{-,\dagger}\!\left(x^+ - v^+, \boldsymbol{k}_1\right)\,	\tilde{\mathcal{A}}_{A,\bara}^{-}\!\left(x^+ - \bvp, \bk_1\right)\, \right\rangle\\
		&\qquad\qquad=\frac{\delta^{a\bara}}{d_A}\left\langle 	\tilde{\mathcal{A}}_{A,a}^{-,\dagger}\!\left(x^+ - v^+, \boldsymbol{k}_1\right)\,	\tilde{\mathcal{A}}_{A,a}^{-}\!\left(x^+ - \bvp, \bk_1\right)\, \right\rangle
	\end{split}
\end{equation}

 Additionally, using that $f_{abc}f^{abc}=N_c d_A$, leads to the color averaged energy-momentum tensor, 
\begin{equation}
	\label{eq:tmunu_fmunu2}
	\begin{split}
		\langle T^{\mu\nu}(x)\rangle=& \;\frac{\as}{C_F}\,
		\int_{\kp_1,\kp_2}\;\int_{\bk_1,\bk_2}  \int_0^\infty \rmd v^+ \int_0^\infty \rmd v^- \int_0^\infty \rmd \bvp\int_0^\infty \rmd \bvm\;t^{\mu\nu}(v, \kp_1,\kp_2;\bar{v}, \bk_1,\bk_2)\,
		\\
		& \times \left\langle 	\tilde{\phi}_A^{a,\dagger}\!\left(x^+ - v^+, \boldsymbol{k}_1\right)\,	\tilde\phi_A^{a}\!\left(x^+ - \bvp, \bk_1\right)\, \right\rangle \left\langle 	\tilde{\phi}_B^{b,\dagger}\!\left(x^- - v^-, \boldsymbol{k}_2\right)\,	\tilde\phi_B^{b}\!\left(x^- - \bvm, \bk_2\right)\, \right\rangle\\ 
		&\times
		\rme^{-\rmi|\kp_1+\kp_2|\,\tau'}\,
		\rme^{\rmi(\kp_1+\kp_2)\cdot\xv}
		\rme^{\rmi|\bk_1+\bk_2|\,\bar{\tau}'}\,
		\rme^{-\rmi(\bk_1+\bk_2)\cdot\xv}
	\end{split}
\end{equation}

At high energy the two nuclei are strongly Lorentz-contracted, so each color
source $\rho_{A,B}(x^\pm,\xv)$ is sharply peaked in its light-cone time and
correspondingly broad in the conjugate light-cone momentum. Throughout this computation we take the eikonal limit, in which the lightcone support of the sources vanishes, 
$\rho_A(x^+,\xv)\to\delta(x^+)\,\rho_A(\xv)$. The corrections from the finite
longitudinal width of the sources, or equivalently, the breaking of boost
invariance, are left for future work.

\subsection{The eikonal limit}
\label{sec:eikonal}
In the Eikonal limit, the sources have only point like support along the light cone, $\mathcal{A}_{A,a}^{-}\!\left(x^+,\xv\right) = \delta(x^+)\mathcal{A}_{A,a}^{-}\!\left( \xv\right)$,
which takes care of all the $v,\bar{v}$ integrations. This leads to the much simpler expression, 
\begin{equation}
	\label{eq:tmunu_fmunu_eik}
	\begin{split}
		\langle T^{\mu\nu}(x)\rangle=& \;\frac{\as }{C_F}\,
		\int_{\kp_1,\kp_2}\;\int_{\bk_1,\bk_2}\;
		\left\langle 	\tilde{\mathcal{A}}_{A,a}^{-,\dagger}\!\left( \boldsymbol{k}_1\right)\,	\tilde{\mathcal{A}}_{A,a}^{-}\!\left( \bk_1\right)\, \right\rangle \left\langle 	\tilde{\mathcal{A}}_{B,b}^{+,\dagger}\!\left( \boldsymbol{k}_2\right)\,	
		\tilde{\mathcal{A}}_{B,b}^{+}\!\left( \bk_2\right)\, \right\rangle	\\
		& \times t^{\mu\nu}(x, \kp_1,\kp_2;x, \bk_1,\bk_2)\,
		\rme^{-\rmi|\kp_1+\kp_2|\,\tau}\,
		\rme^{\rmi(\kp_1+\kp_2)\cdot\xv}
		\rme^{\rmi|\bk_1+\bk_2|\,\tau}\,
		\rme^{-\rmi(\bk_1+\bk_2)\cdot\xv}
	\end{split}
\end{equation}

We will express the correlations of the background fields, in terms of the uGDFs, see \cref{eq:wwfunction}, to write 
\begin{equation}
	\begin{split}
		\left\langle 	\tilde{\mathcal{A}}_{A,a}^{-,\dagger}\!\left( \boldsymbol{k}_1\right)\,
		\tilde{\mathcal{A}}_{A,a}^{-}\!\left( \bk_1\right)\, \right\rangle&= 
		g^2\left\langle     \frac{\tilde{\rho}^\dagger_{A,a}(\kp_1)}{\kp_1^2}\frac{     \tilde{\rho}_{A,a}(\bk_1)}{\bk_1^2}
		\right\rangle\,\\
		=&\;\frac{1}{\pi}   \frac{1} { \kp_1\cdot \bk_1} \int_{\xv} \rme^{-\rmi \xv \cdot (\kp_1-\bk_1) }\varphi_A \left( \frac{\kp_1+\bk_1}{2},\xv\right)
	\end{split}
\end{equation}
And so we get a similar expression to \cref{sec:tmunu-spectrum}, 
\begin{equation}
	\label{eq:tmunu_fmunu_eik_2}
	\begin{split}
		\langle T^{\mu\nu}(x)\rangle=& \;\frac{\as}{C_F\pi^2}\,
		\int_{\kp_1,\kp_2}\int_{\bk_1,\bk_2}\int_{\zv,\yv}\;\frac{t^{\mu\nu}(x, \kp_1,\kp_2;x, \bk_1,\bk_2)}{\left( \kp_1\cdot \bk_1\right)\left( \kp_2\cdot \bk_2\right)}\, \rme^{-\rmi\tau\,(|\kp_1+\kp_2|-|\bk_1+\bk_2|)}\,\\
		&\times\varphi_A \left( \frac{\kp_1+\bk_1}{2},\zv\right)\varphi_B \left( \frac{\kp_2+\bk_2}{2},\yv\right) \rme^{-\rmi (\zv-\xv) (\kp_1-\bk_1) }\rme^{-\rmi (\yv-\xv) (\kp_2-\bk_2) }
	\end{split}
\end{equation}

Notice that in the strict use of the eikonal limit the uGDFs do not have any $x$ dependence, as we have for now a strict eikonal source. Hence, here the uGDFs correspond to a fixed correlation given by the sources. Their $x$-dependence will be restored later when there is a clear kinematic value for the light cone momenta of the outgoing gluons, just as in \cref{sec:tmunu-spectrum}.
In the eikonal limit, the kernel simplifies to the following expression (for the general case see \cref{app:tmunu-reduction}), 
\begin{equation}
	\begin{split}
		t^{+-}&=\frac{h_S\,}{\tau\sqrt{|\Pp||\bar{\Pp}|}},\qquad
		t^{\pm\pm}=\frac{2(x^\pm)^2}{\tau^3(|\Pp||\bar\Pp|)^{3/2}}\big(h_S\,\PdPbar\pm\mathcal D \PXPbar\big),\\
		t^{\pm i}&=\frac{x^\pm}{\tau^2\sqrt{|\Pp||\bar{\Pp}|}}\left[ h_S\!\left(\frac{P^i}{|\Pp|}+\frac{\bar P^i}{|\bar{\Pp}|}\right)
		\mp\mathcal D\!\left(\frac{\tilde P^i}{|\Pp|}-\frac{ \tilde{\bar P}^i}{|\bar{\Pp}|}\right)\right] ,\\[4pt]
		t^{ij}&=\frac{h_S}{\tau \left(|\Pp||\bar{\Pp}|\right)^{3/2}}\bigg[ \left(P^i \bar{P}^j+P^j \bar{P}^i\right)+\delta^{ij} \left(|\Pp||\bar \Pp|-\PdPbar\right)\bigg]
		\label{eq:tmunu_eikonal}
	\end{split}
\end{equation}
where for simplicity of notation, we define $\Pp=\kp_1+\kp_2$ and $\bar \Pp = \bk_1+\bk_2$. Here we keep the $|\Pp|$ instead of $P_\perp$ to highlight the functional dependence of the terms. The physical meaning of such vectors can be understood from the perturbation field-strength tensor definition (or alternatively its field itself, $a^{\mu}$), where $\Pp$ can be understood as a mode momentum. Additionally, we have defined the notation $\tilde P^{i} =\epsilon^{ij}P^j$ Such modes naturally stem from the interaction of gluon modes propagating from the inserted incoming color sources. Additionally,  $h_S$ and $\mathcal{D}$ given by
\begin{equation}
	\begin{split}
		h_S = k_1^i k_2^j  \bar{k}_1^k \bar{k_2}^l (\delta^{ij}\delta^{kl} + \epsilon^{ij}\epsilon^{kl})\quad \text{and} \quad 
		\mathcal{D} = k_1^i k_2^j  \bar{k}_1^k \bar{k_2}^l (\epsilon^{ij}\delta^{kl}-\delta^{ij}\epsilon^{kl} )\\
	\end{split}
\end{equation}
The object $h_S$ is nothing else than the numerator of the hard factor in \cref{eq:hard-factor}, while
$\mathcal{D}$ is its parity-odd partner, the invariant that keeps track of the
mismatch between the amplitude and its conjugate. These two objects scale very
differently in our hierarchy: as a $P$-even object, $h_S$ does not vanish when the
field and its conjugate share the same mode, whereas $\mathcal{D}$ has to be controlled entirely by their mismatch, i.e. it disappears
as $\lp\to0$.

It is interesting to investigate the structure of this kernel  in a more intuitive way using the description of the field-strength tensor in terms of the chromo-electric and chromo-magnetic fields as measured in Milne coordinates. For a general field strength tensor, this is given by
\begin{equation}
	F^{+-}=E_\eta,\qquad
	F^{ij}=\epsilon^{ij}B_\eta,\qquad
	F^{\pm i}=\frac{x^\pm}{\tau}\left(E^i_\perp \pm \tilde B^i_\perp\right),
	\qquad \tilde B^i_\perp\equiv\epsilon^{ij}B^j_\perp\,.
	\label{eq:F_milne_general}
\end{equation}

Comparing \cref{eq:F_milne_general} with the kernel $\tilde f^{\mu\nu}$,  we can read off the
comoving fields of the positive-frequency amplitude. In the large-momentum mode limit, in which the modes can more cleanly be described by plane wave modes,  we can describe the chromo-electric and chromo-magnetic momentum modes as 
\begin{equation}
	E_\eta=\frac{\kp_1\cdot\kp_2}{\sqrt{\tau |\Pp| }},\quad
	B_\eta=\frac{\kp_1\times\kp_2}{\sqrt{\tau |\Pp|}},\quad
	\mathbf E_\perp=\frac{\kp_1\times\kp_2}{\sqrt{\tau |\Pp|}}\,\frac{\tilde{\Pp}}{|\Pp|},\quad\text{and}\quad
	\mathbf B_\perp=-\frac{\kp_1\cdot\kp_2}{\sqrt{\tau |\Pp|}}\,\frac{\tilde{\Pp}}{|\Pp|}\,. 
	\label{eq:milne_dictionary}
\end{equation}

Naturally, this is not supposed to be a formal assignment of the chromoelectromagnetic fields, but more a pedagogical tool to intuitively understand which terms in the computation below, and furthermore in the gradient expansion, arise from which chromo-electromagnetic field components. 
It is straightforward to see that the transverse fields are created as response to the initial longitudinal fields. In the language of this section, this can be seen from $	\mathbf E_\perp=B_\eta \,\tilde{\Pp}/|\Pp|$  and $	\mathbf B_\perp=-E_\eta \,\tilde{\Pp}/|\Pp|$. 
This behavior is expected, as the response of the transverse fields to the initial longitudinal ones has been extensively discussed in the literature~\cite{Chen:2015wia,Lappi:2006fp}.
 Using these expresions, we can appreciate the field-dependence of the expressions, where for example $t^{+- }\sim E_\eta^2 + B_\eta^2$ is related to the longitudinal energy and a term such as $\calD \Pp\times \bar\Pp \sim \mathbf{E}_\perp\times \mathbf{B}_\perp$, corresponding directly to the longitudinal Poynting flux.  While this is by no means surprising, as it is the basic construction of a gaug field energy-momentum tensor,  thinking of the kernel in these terms will be useful later to see understand which terms under the gradient expansion correspond to which physical processes within the energy-momentum tensor.

We can then start the formal gradient expansion. First, we define the following functions to clean the notation,
\begin{equation}
	\begin{split}
		\calH^{\mu\nu}(x; \kp_1,\kp_2;\bk_1,\bk_2) =& \frac{t^{\mu\nu}(x, \kp_1,\kp_2;x, \bk_1,\bk_2)}{\left( \kp_1\cdot \bk_1\right)\left( \kp_2\cdot \bk_2\right)}\,\rme^{-\rmi \tau\, \Phi(\kp_1+\kp_2,\bk_1+\bk_2)}\\
		\text{with }\quad\Phi(\kp_1+\kp_2,\bk_1+\bk_2)=&|\kp_1+\kp_2|-|
		\bk_1+\bk_2|
	\end{split}	 
\end{equation}
We will once again perform a Wigner rotation in the momenta variables, exactly the same way as in \cref{sec:tmunu}, 

\begin{equation}
	\label{eq:wigner_rotation_2_electric_boogaloo}
	\qp\equiv \frac{\kp_1 + \bk_1}{2},
	\qquad
	\kp\equiv\frac{\kp_2 + \bk_2}{2} 
	\qquad
	\tilde{\qp} \equiv \kp_1 - \bk_1,
	\qquad
	\tilde{\kp} \equiv \kp_2 - \bk_2\,,
\end{equation}
In these variables \cref{eq:tmunu_fmunu_eik_2} becomes
\begin{equation}
	\label{eq:tmunu_fmunu_eik_wigner}
	\begin{split}
		\langle T^{\mu\nu}(x)\rangle=& \;\frac{\as}{\pi^2C_F}\,
		\int_{\qp,\kp,\qtil,\ktil}\int_{\zv,\yv}
		\;\calH^{\mu\nu}\left(x; \qp+\frac{\qtil}{2},\kp+\frac{\ktil}{2};\qp-\frac{\qtil}{2},\kp-\frac{\ktil}{2}\right)
		\\
		&\times \varphi_A \left(\qp,\zv\right)\varphi_B \left(\kp,\yv\right)\rme^{-\rmi (\zv-\xv) \cdot\qtil } \rme^{-\rmi (\yv-\xv)\cdot \ktil }
	\end{split}
\end{equation}

We can then perform an expansion on the full kernel analogous to the one performed for the single gluon production formula. In our case now, the kernel is more complex, and it contains already the tensorial information of the energy-momentum tensor.  For this, we define the double Taylor expansion for the relative momenta defines a set of tensor coefficients
\begin{equation}
	\calH^{\mu\nu}\left(x;\boldsymbol{k}_1,\boldsymbol{k}_2;\bar{\boldsymbol{k}}_1,\bar{\boldsymbol{k}}_2\right)=\sum_{n,m=0}^{\infty}\frac{1}{n!\,m!}\tilde{q}^{i_1}...\tilde{q}^{i_n} \tilde{k}^{j_1} ... \tilde{k}^{j_m}\,\calH_{(n,m)}^{\mu\nu;i_1,...,i_n;j_1,...,j_m}(\qp,\kp).
	\label{eq:expansion_position}
\end{equation}
where $\calH_{(n,m)}^{\mu\nu;i_1\dots i_n;j_1\dots j_m}(\qp,\kp)$ represents the $n$-th and $m$-th transverse derivatives of $\calH^\mu$ at $\qtil=\ktil=0$. Once again, it is useful to reorganize the expansion term by grouping all contributions to a certain order $l$, $H=\sum_{l=0}^{\infty}H_l$, where
\begin{equation}
	\calH^{\mu\nu}_l(x,\qp,\kp;\qtil,\ktil)=\sum_{n,m=0}^{\infty}\frac{\delta_{l,n+m}}{n!\,m!}\tilde{q}^{i_1}...\tilde{q}^{i_n} \tilde{k}^{j_1} ... \tilde{k}^{j_m}\,\calH_{(n,m)}^{\mu\nu;i_1,...,i_n;j_1,...,j_m}(\qp,\kp).
	\label{eq:expansion_position_by_order}
\end{equation}

Plugging \cref{eq:expansion_position,eq:expansion_position_by_order} into \cref{eq:tmunu_fmunu_eik_wigner}

\begin{equation}
	\label{eq:tmunu_fmunu_eik_wigner_exp}
	\begin{split}
		\langle T^{\mu\nu}(x)\rangle=& \;\frac{\as}{\pi^2C_F}\,\,\sum_{l=0}^{\infty}
		\int_{\qp,\kp,\qtil,\ktil}\int_{\zv,\yv}
		\;\calH^{\mu\nu}_l(x,\qp,\kp;\qtil,\ktil)
		\\
		&\times \varphi_A \left(\qp,\zv\right)\varphi_B \left(\kp,\yv\right)\rme^{-\rmi (\zv-\xv) \cdot\qtil } \rme^{-\rmi (\yv-\xv)\cdot \ktil }
	\end{split}
\end{equation}

We will play the same game as in the last section, where we will switch the $\qtil, \ktil$ dependence to the transverse gradients. This will result in the exponentials carrying the only vestigial dependence on the relative momenta, which can be integrated out to get deltas. After all the integrations we get,

\begin{equation}
	\label{eq:tmunu_fmunu_eik_wigner_nabla}
	\begin{split}
		\langle T^{\mu\nu}(x)\rangle=& \;\frac{\as}{\pi^2 C_F}\,\sum_{l=0}^{\infty}
		\int_{\qp,\kp} \bigg[\calH_l^{\mu\nu}(x,\qp,\kp;-\rmi\nabla_{\zv},-\rmi\nabla_{\yv}) \varphi_A \left(\qp,\zv\right)\varphi_B \left(\kp,\yv\right)\bigg]_{
			\yv=\zv=\xv}\\
		&=\sum_{l=0}^{\infty}\langle T_{(l)}^{\mu\nu}\rangle.	
	\end{split}
\end{equation}
To show that this framework contains the information in \cref{sec:tmunu-spectrum} plus extra information related to infrared field-like dynamics, we will expand the same way as before, to first non-trivial order ($l=2$).

\subsection{Expansion around the ``on-shell'' limit}
\label{sec:on-shell-expansion}

As stated in the introduction and in the last section, three transverse scales control the dynamics of the evolution. First, the softest scale is the inverse transverse size of the sources, $1/R$. This is the scale  over which the color-charge densities, and eventually the
saturation scales $Q_s^2(\xv)$, defined from the former, vary appreciably. 

The intermediate scale is given by the saturation scales $Q_{s}$ within the targets. These scales set the typical internal momentum within each target and about which the unintegrated distributions
$\varphi_{A,B}(\qp,\xv)$ are peaked.  Furthermore, in the dilute limit, the mother gluons (the gluons probed from the uGDFs with average momenta $\qp$ and $\kp$, in the momentum-space computation language), are expected to be larger than $Q_s$. For this reason, the hardest scale in the dilute-dilute limit is the momentum $|\pp|$ of the produced
gluonic mode, which we take to be much harder than the saturation scale, $p_\perp\gg Q_s$. Summarizing, $R^{-1} < Q_s \lesssim q_\perp, k_\perp < |\pp|$.
In other words, the mean and relative
momenta of the Wigner decomposition inherit the same ordering. The average momenta $\qp,\kp\gtrsim Q_s$
probe the internal structure of the sources, while the relative momenta
$\tilde\qp,\tilde\kp$  
measure the field--conjugate mismatch,  order of the size of the source inhomogeneity.

 Applying the Wigner rotation in \cref{eq:wigner_rotation_2_electric_boogaloo} to these variables gives us $\Pp = \pp + \lp/2$ and $\bar \Pp = \pp - \lp/2$, where  $\pp=\qp+\kp$ and $\lp=\qtil+\ktil$. 
This is important since $\Pp$ and $\bar \Pp$ appear within the energy-momentum  kernel $t^{\mu\nu}$, \cref{eq:tmunu_eikonal}. 
Based on the hierarchy of scales in the problem, $\Pp=\pp+\lp/2$ and $\bar\Pp=\pp-\lp/2$, expand every factor that
depends only on $\Pp,\bar\Pp$ in powers of $\lp/|\pp|$. The leading terms scale
like $(R|\pp|)^{-2}$, see for example, 
\begin{equation}
	\begin{split}
	\label{eq:norm_expansion}
	\frac{1}{\sqrt{|\Pp||\bar{\Pp}|}} = \frac{1}{|\pp|} \left[1+\frac{ \left(2 (\pp.\lp)^2-\pp^2 \lp^2\right)}{8 |\pp|^4}+\calO\left((R|\pp|)^{-4}\right)\right] 
	\end{split} 
\end{equation}
while the scalar and cross products give exactly 
\begin{equation}
	\Pp \cdot \bar{\Pp} = \pp^2 -\frac{1}{4}\lp^2 \quad \text{and} \quad \Pp \times \bar{\Pp}= -\pp \times \lp
\end{equation}
Contrasting this is the denominator of $H^{\mu\nu}$, which contains terms like
$1/(\kp_1\cdot\bk_1)$. Due to the fact that such a series depends on powers of $\qp$ and $\kp$ individually, this term in principle expands in a separate series, whose leading term scales
like $1/\qp^2$ with corrections of order $1/(Q_s R)$. These terms, as we will see in the following are less suppressed than the ones suppressed by powers $1/(p_\perp R)$ when the scale hierarchy is enforced. 

For this reason we can  split the calculation into two
different expansions. Expanding first the terms that depend uniquely on $\Pp$ and
$\bar\Pp$ to first non-trivial order,  ($\calO(\lp^2)$, $l=2$ in last sections notation), what we get is that the energy-momentum tensor kernel splits into three contributions\footnote{The second non-trivial order ($l=4$) is quite involved and will be computed in future work. }, 
\begin{equation}
	t^{\mu\nu} = t^{\mu\nu}_{os} +t^{\mu\nu}_\epsilon + \delta t^{\mu\nu} 
\end{equation}
where $t^{\mu\nu}_{os}$  is the energy-momentum tensor in the on-shell limit, the
limit in which the amplitude and its conjugate amplitude share the same mode
momenta, just as when computing the gluon yields. The on-shell kernel simplifies to the expression
\begin{equation}
	t^{\mu\nu}_{os}=\frac{2\,h_S\,}{\tau \,|\pp|^{3}}\;P^{\mu}P^{\nu},
	\qquad
	P^{\mu}=\Big(\tfrac{|\pp|}{\sqrt{2}} \rme^\eta,\;\tfrac{|\pp|}{\sqrt{2}}\rme^{-\eta},\;\pp\Big)
\end{equation}
from which it is straightforward to show that one recovers the main results from \cref{sec:tmunu}. Intuitively, this on-shell term stands for a group of modes transporting the longitudinal energy $\sim h_S$ forward. Notice that here we have included the kinematic factor $h_S$ completely, instead of order by order in the gradients. This is to show that using an expansion around a coherent final state $\Pp = \bar{\Pp} = \pp$ one can recover the complete yield from the momentum-space computation, and single it out as the largest contribution order by order, using the problem's scale hierarchy~\footnote{In accompanying phenomenological paper we will show this holds for a variety of systems relevant to the LHC. }. Nevertheless, when expanding around $\qtil,\ktil, \lp$ to a fixed order, or equivalently the gradients, one must match first order by order in the gradients, and then compare the remaining scalings to decide  which orders are worth keeping. 

Apart from the on-shell contributions, after the expansion around $\lp$, we get a piece that simplifies into a simplified form 
\begin{equation}
	\tau t ^{\mu\nu}_\epsilon=\frac{p^\mu p^\nu h_S^{(0)}}{4|\pp|^7}\left(2 (\pp\cdot \lp)^2 - \lp^2\pp^2\right) = \frac{p^\mu p^\nu  }{4|\pp|^5} \qp^2\kp^2\lp^2 \cos\left(2\phi_{\lp \pp}\right)\,
	\label{eq:tmunu_epsilon}
\end{equation}

When translating this operator into gradient form $\lp\to-\rmi(\nabla_{\xv}+\nabla_{\yv})$, \cref{eq:tmunu_epsilon} gives   $-(2\hat p^i\hat p^j-\delta^{ij})\,\nabla^i\nabla^j[\varphi_A\varphi_B]$. This traceless transverse Hessian operator (over the local uGDF overlap) results in enhanced emission along fastest descent of the overlap, and is  depleted orthogonal to it. In other words, the outgoing gluon momentum tries to align itself onto the density-gradient axis, ultimately deforming the emission pattern.
As we will see later in this section, this is distinct, to \cref{eq:LOpretty}, since this term can be rewritten explicitly as a response to the local zeroth-order gluon density, while \cref{eq:LOpretty}, contained in the on-shell contribution, measures a different moment of the  "overlap distribution. in fact, this can be explained by noticing that this term is nothing but the first correction in 
\cref{eq:norm_expansion}. Finally, this term scales with an extra power of $(R|\pp|)^{-2}$ compared to the zeroth order contribution, the {\ktfact} formula. If one would compare it to the contribution in \cref{eq:LOpretty}, one would get an extra suppression of $(Q_s/p_\perp)^2$, which means that this term will be numerically suppressed with respect to the terms in $t_{os}^{\mu\nu}$.


In contrast, the rest of the corrections of the energy-momentum tensor kernel at this level, parametrized by $\delta t^{\mu\nu}$, 
 do not simplify into a compact form like the on-shell case or the elliptical response. However, the cumbersome form of this contribution presents itself a bit better in Milne coordinates, where it is easier to understand the origin of the terms. The   $\delta t^{\mu\nu}$ object is given componentwise by
\begin{equation}
	\label{eq:deltatmunu}
	\begin{split}
		\tau \delta t^{\tau\tau}&=-\frac{h_S^{(0)}}{2|\pp|^5} (\lp\times \pp)^2\,, \quad 
		\tau \delta t^{\tau\eta}= -\frac{1 }{\tau |\pp|^3}\calD_1 \, (\pp\times \lp ) \\
		\tau \delta t^{\tau i }&= -\frac{p^i }{4 |\pp|^6} h_S^{(0)}\left( 3(\lp\cdot \pp)^2 - \lp^2 \pp^2\right)   \,, \quad
		\tau \delta t^{\eta i }= -\frac{p^i }{\tau |\pp|^4}\calD_1 \, (\pp\times \lp ) \\
		\tau \delta t^{ij}&= \frac{ h_S^{(0)}}{2|\pp|^7}(\lp\cdot \pp)(\lp \times\pp)(p^i  \tilde{p}^j+p^j\tilde{p}^i ) \quad \text{and} \quad \tau \delta t^{\eta\eta}=\frac{1}{\tau}  \delta t^{\tau\tau}
	\end{split}
\end{equation}
where $h_S^{(0)}=\qp^2\kp^2$ is the zeroth order term in the gradient expansion of $h_S$. It follows that then $\calD_1$ is the first order expansion term of $\calD$, with
\begin{equation}
	\calD_1= \delta^{ij}\epsilon^{kl} \left[ q^ik^j\left(q^k\tilde{k}^l + \tilde{q}^k k^l	\right)  - q^kk^l\left(q^i\tilde{k}^j + \tilde{q}^i k^j	\right)  \right]
\end{equation}

At first glance the components split into two groups, set by the two invariants of the
longitudinal color fields. The components $\delta t^{\tau\tau}$, $\delta t^{\tau i}$, $\delta t^{ij}$ and
$\delta t^{\eta\eta}$, symmetric to longitudinal parity transformations are driven by the  coefficient $h_S^{(0)}$, the 
zeroth-order of the deposited energy of the longitudinal fields,
$h_S\leftrightarrow |E^\eta|^2+|B^\eta|^2$. Because of this, these components describe how that deposited energy is redistributed in the
transverse plane via the gradients of the overlap. First we have $\delta t^{\tau\tau}$, which is a normalization correction of the energy density, operating through the curvature of the overlapping regions transverse to the mode momentum 
$\pp$, $(\lp\times\pp)^2=\lp^2\pp^2-(\lp\cdot\pp)^2$. The first correction to the transverse Poynting vector is given by  $\delta t^{\tau i}$ and through the $\big(3(\lp\cdot\pp)^2-\lp^2\pp^2\big)$ term, which is a pre-flow set by the response of the
transverse fields to gradients of the longitudinal ones~\cite{Chen:2015wia}. Operationally, since $(\lp\cdot\pp)^2\propto\cos^2(\phi_l-\phi_p)$, the $\big(3(\lp\cdot\pp)^2-\lp^2\pp^2\big)$ structure combine total curvature
 ($\nabla^2$) piece with a quadrupole modulation $\propto\cos 2(\phi_l-\phi_p)$. Finally, we get the $\delta t^{ij}$ term, which is a pure shear stress term modulated by a quadrupole sine function. Together with the elliptic piece \cref{eq:tmunu_epsilon}, it spans the full transverse anisotropic
pressure at this order. The longitudinal component $\delta t^{\eta\eta}=\delta t^{\tau\tau}/\tau^2$ is constrained by
the tracelessness of $T^{\mu\nu}$.

On the other hand, we have the $\eta$-odd components $\delta t^{\tau\eta}$ and $\delta t^{\eta i}$ driven by the coefficient $\calD_1$, the leading gradient term of 
$\calD\leftrightarrow\langle B^\eta\bar E^\eta-E^\eta\bar B^\eta\rangle$. 
The 
$\delta t^{\tau\eta}$ component is the longitudinal Poynting component
$\sim\mathbf E_\perp\times\mathbf B_\perp$, and $\delta t^{\eta i}$ its transverse--longitudinal
shear partner. It is interesting to see that since these structures do not contain $p^\eta$, they do not trivially vanish, as opposed to the kinetic terms, signaling that we are now in field dynamics territory. Physically, this is a coherent longitudinal flow, built from the interference of the two nuclei's longitudinal fields. It is interesting that this term, even though it is suppressed, arises from the transverse inhomogeneities alone, even within the eikonal limit. Seeing that in Ref.~\cite{Ipp:2024ykh} the rapidity-odd flow comes from the longitudinal extent of the sources, it would be really interesting to contrast this term-by-term expansion against the eikonal limit.

Once again, we want to stress that  order-by-order in the relative momenta, and hence in the  gradients, the elliptical and $\delta t$ corrections are higher orders, adding extra powers of $(Q_s^2/\pp^2)$. In the dilute-dilute limit scale hierarchy, these terms are numerically suppressed.

Finally we come to the phase term. We will continue the same line of argument and
expand it using the hierarchy between $\pp$ and $\lp$, to get
\begin{equation}
	\Phi\left(\pp+\frac{1}{2}\lp, \pp-\frac{1}{2} \lp\right)=\left|\pp+\frac{\lp}{2}\right|-
	\left|\pp-\frac{\lp}{2}\right|=\frac{\pp\cdot\lp}{|\pp|}
	\left[1+ \calO\!\left(\left(|\pp| R\right)^{-2}\right)\right]
\end{equation}
What is interesting here is that, due to the self-normalization of the total
momentum $\pp$, the phase now scales as $\tau\lp\sim\tau/R$, instead of the naive
$\tau|\pp|$, and each correction term carries extra powers of $(|\pp|R)^{-2}$. This scaling is what ensures the validity of the expansion, giving us
a condition to satisfy with the selected scales at the level of computing
observables. For early-time applications of this formula, this means that as long
as the source size is larger than the time of initialization, we can trust this
approximation. To keep consistency with the above expansion we will keep only terms up to second order in $\lp$, which means the phase term results in  
	\begin{equation}
		\rme^{-\rmi \tau\Phi\left(\pp+\frac{1}{2}\lp, \pp-\frac{1}{2} \lp\right)}=1-\rmi \tau\frac{\pp\cdot\lp}{|\pp|} \to 1- \tau\frac{1}{|\pp|} \pp\cdot\left(\nabla_{\zv} +\nabla_{\yv}\right) 
	\end{equation}
What is worth noting is that the phase introduces a transverse,
$P$-odd term that does not scale with $\pp$. This comes in contrast with the
$P$-odd terms in the full $t^{\mu\nu}$, such as $\mathcal{D}$ and
$\pp\times\bar\pp$, which always appear with extra suppressing powers of the
outgoing momentum $\pp$. The ulterior consequence is that the phase introduces 
$P$-breaking even in the particle limit, allowing for a momentum flow in the
system which is forbidden in the momentum computation. It is easy to see this by looking at a component of the on-shell leading order contribution such as $t_{os}^{+i}$ (or $t_{os}^{\tau i}$ if in Milne coordinates). The extra $\pp$ insertion in the second term of the approximated phase shifts $t_{os}^{+i}$ towards a non-vanishing value. The result is a first-order gradient term, arising entirely from the interference phase between the amplitude and its conjugate.

Up to first non-trivial order, i.e. second order in gradients, we have then the energy-momentum tensor
\begin{equation}
	 \expval{T^{\mu\nu}_{\phantom{(T)}} }= \expval{T^{\mu\nu} _{(0)}}+
	\expval{T^{\mu\nu}_{(1)}} +
\expval{T^{\mu\nu}_{(2)}} 
		\end{equation}
where $\expval{T^{\mu\nu} _{(0)}}$ is the {\ktfact} formula \cref{eq:ktfactorization}. This zeroth-order term contains only the kinetic part where the energy
density and isotropic pressure of the on-shell modes is sensitive to the field solely
through the magnitude of the mode energy, $\mathbf E^2+\mathbf B^2$. The first order correction arises due to the first order term from the phase, and it is given by 
\begin{equation}
	\expval{T^{\mu\nu}_{(1)}}= - \int \rmd^2 \pp \left[\frac{p^\mu p^\nu}{p^\tau}\hat{\pp}\cdot\nabla_{\xv} \left(\frac{\rmd N^{(0)}}{\rmd^2 \xv \rmd^2 \pp \rmd y}\right)\right]_{\eta=y}
\end{equation}
where $\hat{\pp}= \pp/|\pp|$. Due to the vectorial content of this integral, integrating over the full domain of $\pp$ renders $	\expval{T^{\mu\nu}_{(1)}}$ non-vanishing exclusively for the $	\expval{T^{\tau i}_{(1)}}$ and $	\expval{T^{\eta i}_{(1)}}$ components. It is important to note that for isotropic distributions one can integrate the angular variable to find $ -\nabla_{\xv}^i T^{\tau\tau}$, which sets the transverse velocity fields to switch on as the response to gradients of the initial deposited energy. This result is similar to earlier works \cite{Chen:2013ksa,Chen:2015wia} where the response was computed in the small-$\tau$ approximation. 
The appearance of a momentum current term, which is the linear response to an initial local densities is one of the main results of this work.

The second order energy-momentum tensor is slightly more complex, but we can split it using the separation of the kernel used above. In this case, we define, using analogous notation
\begin{equation}
		\expval{T^{\mu\nu}_{(2)}} = \expval{T^{\mu\nu}_{(2,os)}} + \expval{T^{\mu\nu}_{(2,\epsilon)}} + \expval{\delta T^{\mu\nu}_{(2)}}
\end{equation}
where the first term is exactly equivalent to the second order contribution in \cref{eq:Tmunu_os_final}. The elliptical piece is given within the complete formula by 
\begin{equation}
	\tau \expval{T^{\mu\nu}_{(2,\epsilon)}} =- \frac{1}{8}\int \rmd^2 \pp \left[\frac{p^\mu p^\nu}{p^\tau}\frac{1}{|\pp|^6}\left(2p^ip^j - \delta^{ij} \pp^2\right)\nabla_{\xv}^i\nabla_{\xv}^j \left(\frac{\rmd N^{(0)}}{\rmd^2 \xv \rmd^2 \pp \rmd y}\right)\right]_{\eta=y}
	\label{eq:tmunu_2_epsilon}
\end{equation}
This part  adds the traceless symmetric part of the stress, i.e.\ the anisotropic (quadrupolar) shear which stems from the transverse chromoelectric and chromomagnetic fields via $E^iE^j+B^iB^j$. 

The last piece unfortunately does not factorize in such a compact form, and one has to use the complete integrand. The remaining piece of the contribution up to second order in gradients is given by 
\begin{equation}
	\begin{split}
		\expval{\delta T^{\mu\nu}_{(2)}}=\frac{\as}{\pi^2 C_F}\,
		\int_{\pp,\qp,\kp} (2\pi)^2 \delta^{(2)} (\pp-\qp-\kp) \bigg[\delta t^{\mu\nu}(x,\qp,\kp;-\rmi\nabla_{\zv},-\rmi\nabla_{\yv}) \\ 
		\times \frac{\varphi_A \left(x_A, \qp,\xv\right)}{\qp^2} \frac{\varphi_B \left(x_B, \kp,\yv\right)}{\kp^2}\bigg]_{
			\yv=\xv}
		\end{split}
	\label{eq:d_tmunu_second_order}
\end{equation}

In fact, the structure of \cref{eq:d_tmunu_second_order} is most transparent if one reads the
gradient expansion as a linking term within a controlled interpolation between a gas of on-shell gluons and
the underlying classical field.
The growing
infrared sensitivity of this terms is a  telling sign of this scheme. That is why we can think of $\delta T^{\mu\nu}$ as field-like corrections
which are both collective and soft-mode dominated. Moreover, this term encodes the first non-kinetic level of the information discarded by the on-shell projection.

A final note on the infrared dominance. By quick inspection it is easy to see that formulas \cref{eq:tmunu_2_epsilon,eq:d_tmunu_second_order} seem to be infrared divergent on the $\pp$ integration. Hence, a regularization is needed. In the accompanying paper \cite{Garcia-Montero:2026NextPaper}, the explicit regularization and regulator dependence will be explored. The impact of these higher order terms, including their parameter dependence,  will be presented pedagogically using a phenomenologically useful, but simple model. The result is that in practice, the hierarchy presented here is satisfied when computing phenomenologically relevant quantities.


\section{Summary and conclusions}
\label{sec:summary}

In this work we have computed a systematic gradient-expansion  of the energy-momentum tensor within the dilute--dilute computation of single-inclusive gluon production in the CGC. These gradients arise from interpreting the average of the amplitude and conjugate momenta in the observables as the transverse momentum of the
	produced gluon, while understanding that their difference is the conjugate variable to the impact parameter at
	which the uGDFs are probed. In this way, each power of the relative momentum acts as a
	transverse gradient of the smooth unintegrated distribution. We have shown that  basic  expansion
	parameter is $1/(Q_s R)$, with $R$ the scale over which the sources vary. 
	
	 We have  contrasted this expansion in two different methods which have 
	been used to compute $T^{
	\mu\nu}$ in this limit before. On one hand, we have started from the gluon spectrum, and taking it as a collection of particles, we have computed the energy-momentum tensor by identifying the spectrum with a one-particle distribution. 
	Using this method, we have recovered the 
	standard \ktfact{} formula for
	single-inclusive gluon production, which is the zeroth order of this expansion and the only non-vanishing term in the limit of an infinite source. We
		computed the tower up to the second non-trivial order and showed that the expansion 
	extends the $k_\perp$-factorized formula, retaining the spatial information of the sources without  effectively breaking the factorization.
	
    Up to fourth order in the gradients, we find that the corrections are dominated by a tensor structure which factors the gradient dependence into two different effects. On one side, an isotropic rescaling of the local yield, set by how the two overlapping
profiles vary together. The anisotropic term, instead couples the density gradients to the outgoing gluon
momentum, which effectively picks a preferred transverse direction and produces a pressure
imbalance $T^{xx}-T^{yy}\neq0$ together with a shear $T^{xy}$, already at the level of the gluon yields. This means that these anisotropies are within the medium before any hydrodynamic response.

We then repeated the same computation directly in position space, starting from
the classical Glasma field strength of the dilute limit~\cite{Ipp:2021lwz,Ipp:2024ykh}.
Due to genuine field interference, the modes in the field and conjugate fields are not set to the same momentum. However, we showed that within the hierarchy of scales in the dilute-dilute limit, one can expand around a coherent mode projection, that we have called the on-shell limit.
In this limit, we recover completely the results from \cref{sec:tmunu}. We find higher orders corrections
away from the on-shell limit, where the field kernel splits into an elliptic response (which rotates the emission
pattern onto the density-gradient axis), and a remainder $\delta t^{\mu\nu}$ which
no longer factorizes and carries the parity-odd chromo-electric--magnetic
field pieces. This part of the expansion is controlled by a second ratio,
$1/(p_\perp R)$. In particular, the interference phase between the field and its
conjugate introduces a transverse term which, unlike the
higher orders within the energy-momentum kernel, is not suppressed by the outgoing momentum. This 
generates a first-order momentum current, the linear response of the transverse
flow to gradients of the deposited energy. We report that  similar results were obtained earlier from a small proper-time expansion of the
Glasma~\cite{Chen:2013ksa,Chen:2015wia} where the gradient expansion came from breaking
the factorization of momentum-impact parameter within the individual nuclei. On the other hand, in this work they come from the
real-time interaction of the two colliding fields.

The message that emerges is that the local isotropy of the transverse stress in
the standard $k_\perp$-factorized initial condition is an artifact of the leading
approximation, and not a physical statement: it drops the transverse gradients of
the two finite nuclei by construction. Once these are kept, a momentum-space
anisotropy is already present in the single-gluon spectrum. 
Importantly, such anisotropies are a property of the dynamical mechanism and do not depend on the model
chosen for the unintegrated distributions, although quantitatively will affect their actual value. the $k_\perp$-factorization ansatz only
enters to keep the corrections finite at small momenta. It is also worth recalling that this mechanism is also different from the mechanism behind the glasma-graph
correlations, which generate azimuthal structure from two-point correlators even
in a homogeneous medium. In our computation, we obtain instead anisotropies already at the level of the single-inclusive gluon spectrum, and these are
tied to the collision geometry. So, pictorially, this dynamical mechanism serves as a sort of interpolation between the geometrical scale and the microscopic dynamics. 

Because the effect is
controlled by the variation of the density profiles, it will be largest in compact,
fluctuation-dominated systems. For this reason the natural place to look for it is
the light-ion program at the LHC, O+O and Ne+Ne~\cite{Garcia-Montero:2026oonene}.
The inclusion of initial anisotropies and flow may in fact affect how small
systems approach a hydrodynamic regime~\cite{Garcia-Montero:2026Salgado}, and how
much of the measured $v_n$ is genuinely a hydrodynamic
response~\cite{Ambrus:2022koq,Ambrus:2022qya,Ambrus:2024eqa}. In general terms, it is important to keep in mind that these initial anisotropies will affect the efforts of the community to interface with low-energy nuclear structure, where information about the nuclear  two-body density is proposed to be extracted from 
observables~\cite{Jia:2022ozr,CMS:2025tga}. For this reason, an initial-state anisotropy, largest for light-ion systems, that is not purely a hydrodynamics resoponse, feeds
directly into that program.

A useful feature of the formulation is that every gradient correction is a local
operator acting on the same unintegrated distributions as the zeroth-order
formula, so it can be added directly to a saturation-based initial-state model
such as the \Dipper{}~\cite{Garcia-Montero:2023gex,garcia_montero_2026_20179552},
or written as a cheaper TRENTo-like parametric formula. By keeping the $x$ dependence of the uGDFs, the corrections also keep
a rapidity dependence as in the  factorized formula, so the anisotropies are represented in the whole 3D extension of the events. The latter
phenomenology is developed in the accompanying
paper~\cite{Garcia-Montero:2026NextPaper}. Implemented in an event generator,
these corrections would also refine the ongoing comparison of saturation-based
initial states against other dynamical initialization frameworks~\cite{Constantin:2026hwh}.

It would be interesting to extend this calculation in several directions within the position space computations. The most exciting direction would be of course to relax the eikonal approximation. By expanding in terms of moments of the non-eikonal support of the nuclei, one would be able not only to create correction formulas, as in this work, but also to identify order by order the nature of the boost-invariance breaking terms. Another possible application would be to perform an analogous computation for conserved-current production, through valence
quarks \cite{  McLerran:2018avb,Dumitru:2002qt} and more interestingly the NLO $q\bar q$ channel from gluon fusion~\cite{Gelis:2003vh},
where the same mechanism should show up in charge correlations. On a 
phenomenological side, an initial-state momentum flow that is not of hydrodynamic
origin should be visible in probes sensitive to the earliest stages of the
collision, such as dilepton polarization~\cite{Wu:2024vyc,Coquet:2023wjk} and
direct-photon flow~\cite{Wu:2024pba,Garcia-Montero:2023lrd,Garcia-Montero:2024lbl}.


\subsection*{Acknowledgments}

I would like to thank Nestor Armesto, Carlos Salgado, Bin Wu, Pablo Guerrero-Rodríguez, Sören Schlichting and Travis Dore for valuable discussions. I acknowledge the use of Anthropic's Claude (Claude Code) for assistance cleaning the main text. I have been supported by the European Research Council under project ERC-2018-ADG-835105 YoctoLHC, and by Maria de Maeztu excellence unit grant CEX2023-001318-M.

\appendix


\section{Limits of the tensor kernel $t^{\mu\nu}$}
\label{app:tmunu-reduction}

I will collect in this appendix the closed forms of the kernel
$t^{\mu\nu}(v,\kp_1,\kp_2;\bar v,\bk_1,\bk_2)$ defined in \cref{eq:tmunu}, using the components of \cref{eq:fieldstrength_large}.  The Levi-Civita symbol in transverse 2D is given by $\epsilon^{ij}$. In what follows, 
I use the total transverse momenta and the two rotational invariants of each pair,
\begin{equation}
	\Pp=\kp_1+\kp_2\,,\qquad
	\tilde P^i\equiv\epsilon^{ij}P^j\quad \text{and}\quad S=\kp_1\!\cdot\!\kp_2,\quad A=\epsilon^{ij}k_1^i k_2^j\equiv\kp_1\!\times\!\kp_2,
\end{equation}
with their barred equivalent, e.g. $\bar S$, by applying $\kp_{i}\to \bk_i$
together with the two bilinear invariants and the two overlaps
\begin{equation}
	h_S\equiv A\bar A+S\bar S,\qquad
	\mathcal D\equiv A\bar S-S\bar A
\end{equation}
We abbreviate $L=|\Pp|\,\tau'$ and $\bar L=|\bar{\Pp}|\,\bar\tau'$, and
define the two transverse momentum "currents" objects, given
\begin{equation}
	\boldsymbol{\mathcal U}=A\,\tilde{\Pp}-S\,\Pp,\qquad
	\boldsymbol{\mathcal V}=A\,\tilde{\Pp}+S\,\Pp,
\end{equation}
with $\bar{\boldsymbol{\mathcal U}},\bar{\boldsymbol{\mathcal V}}$ the barred analogues. 
The color-traced and symmetrized kernel is
\begin{equation}
	t^{\mu\nu}
	=\tilde f^{\mu\rho}\,\bar{\tilde f}_{\rho}{}^{\nu}
	+\tilde f^{\nu\rho}\,\bar{\tilde f}_{\rho}{}^{\mu}
	+\tfrac12\,g^{\mu\nu}\,\tilde f^{\rho\sigma}\bar{\tilde f}_{\rho\sigma},
	\qquad
	\bar{\tilde f}\equiv\tilde f(\bar v,\bk_1,\bk_2),
\end{equation}
where the field-strength components, written in the above notation are given by 
\begin{equation}
	\tilde f^{+-}=\frac{S}{\sqrt{L}},\quad
	\tilde f^{ij}=\frac{\epsilon^{ij}A}{\sqrt{L}},\quad
	\tilde f^{+i}=-\frac{v^+}{L^{3/2}}\,\mathcal U^i,\quad
	\tilde f^{-i}=-\frac{v^-}{L^{3/2}}\,\mathcal V^i .
\end{equation}

\subsection*{General case}

A direct contraction and a bit of algebra gives, for the independent components,
\begin{equation}
	\begin{split}
			t^{+-}&=\frac{h_S\,}{\sqrt{L\bar L}},\\[4pt]
		t^{++}&=\frac{2 v^+\bar v^+}{(L\bar L)^{3/2}}\big(h_S\,\PdPbar+\mathcal D \PXPbar\big),\qquad
		t^{--}=\frac{2 v^-\bar v^-}{(L\bar L)^{3/2}}\big(h_S\,\PdPbar-\mathcal D \PXPbar\big),\\[4pt]
		t^{+i}&=\frac{1}{\sqrt{L\bar L}}\left[ h_S\!\left(\frac{v^+P^i}{L}+\frac{\bar v^+\bar P^i}{\bar L}\right)
		-\mathcal D\!\left(\frac{v^+\tilde P^i}{L}-\frac{\bar v^+\tilde{\bar P}^i}{\bar L}\right)\right] ,\\[4pt]
		t^{-i}&=\frac{1}{\sqrt{L\bar L}}\left[ h_S\!\left(\frac{v^-P^i}{L}+\frac{\bar v^-\bar P^i}{\bar L}\right)
		+\mathcal D\!\left(\frac{v^-\tilde P^i}{L}-\frac{\bar v^-\tilde{\bar P}^i}{\bar L}\right)\right] ,\\[4pt]
		t^{ij}&=-\frac{v^+\bar v^-}{(L\bar L)^{3/2}}\big(\mathcal U^i\bar{\mathcal V}^j+\mathcal U^j\bar{\mathcal V}^i\big)
		-\frac{v^-\bar v^+}{(L\bar L)^{3/2}}\big(\mathcal V^i\bar{\mathcal U}^j+\mathcal V^j\bar{\mathcal U}^i\big)
		+\frac{2 A\bar A}{\sqrt{L\bar L}}\,\delta^{ij}
		-\frac12\,\delta^{ij}\,\mathfrak{S},
	\end{split}
\end{equation}

By using these results and performing some algebraic reductions, one gets the scalar component of the energy-momentum tensor kernel, 
\begin{equation}
\begin{split}
		\mathfrak{S}\equiv\tilde f^{\rho\sigma}\bar{\tilde f}_{\rho\sigma}
	=&\frac{2}{\sqrt{L\bar L}}\bigg[(A\bar A-S\bar S)\left(1-\frac{(v^+\bar v^-+v^-\bar v^+)\,\Pp\cdot\bar{\Pp}}{L\bar L}\right)\\ &\qquad\qquad 
	-\frac{(A\bar S+S\bar A)\,(v^-\bar v^+-v^+\bar v^-)\,\PXPbar}{L\bar L}\bigg].
\end{split}
\end{equation}
The whole tensor is thus controlled by the single ``aligned'' invariant $h_S\,=A\bar A+S\bar S$
(the same combination that appears in the hard factor of the single-gluon yield,
\cref{eq:hard-factor-contracted}) together with the ``rotated'' invariant $\mathcal D=A\bar S-S\bar A$,
which is antisymmetric under $(\,\kp_i\!\leftrightarrow\!\bk_i,\,v\!\leftrightarrow\!\bar v\,)$.

\subsection*{Eikonal limit}

In the eikonal limit the light-cone times coincide, $v^\pm=\bar v^\pm=x^\pm$ and
$\tau'=\bar\tau'=\tau$, so $L=|\Pp|\tau$, $\bar L=|\bar{\Pp}|\tau$ and
$v^-\bar v^+-v^+\bar v^-=0$. 
The scalar contraction reduces to
\begin{equation}
	\mathfrak{S}=\frac{2}{\tau\sqrt{|\Pp||\bar{\Pp}|}}(A\bar A-S\bar S)\left(1-\frac{\PdPbar}{|\Pp||\bar\Pp|}\right)
\end{equation}
The independent components are simplified now to 
\begin{equation}
	\begin{split}
			t^{+-}&=\frac{h_S\,}{\tau\sqrt{|\Pp||\bar{\Pp}|}},\qquad
		t^{\pm\pm}=\frac{2(x^\pm)^2}{\tau^3(|\Pp||\bar\Pp|)^{3/2}}\big(h_S\,\PdPbar\pm\mathcal D \PXPbar\big),\\
		t^{\pm i}&=\frac{x^\pm}{\tau^2\sqrt{|\Pp||\bar{\Pp}|}}\left[ h_S\!\left(\frac{P^i}{|\Pp|}+\frac{\bar P^i}{|\bar{\Pp}|}\right)
		\mp\mathcal D\!\left(\frac{\tilde P^i}{|\Pp|}-\frac{ \tilde{\bar P}^i}{|\bar{\Pp}|}\right)\right] ,\\[4pt]
		t^{ij}&=\frac{h_S}{\tau \left(|\Pp||\bar{\Pp}|\right)^{3/2}}\bigg[ \left(P^i \bar{P}^j+P^j \bar{P}^i\right)+\delta^{ij} \left(|\Pp||\bar \Pp|-\PdPbar\right)\bigg]
	\end{split}
	\end{equation}
By using the definition of the light-cone variable $x^\pm = \tau \rme^{\pm \eta}/\sqrt{2}$ it is easy to see that the total dependence of the $t^{\mu\nu}$ kernel in the eikonal limit is $t^{\mu\nu}\sim \tau^{-1}$.
\subsection*{Eikonal limit at coincident total momentum, $\Pp=\bar{\Pp}=\pp$}

Imposing in addition $\Pp=\bar{\Pp}=\pp$ (as enforced, e.g., by the
transverse-momentum conservation of the yield, \cref{eq:momentumconservation}) gives $\PdPbar=\pp^2$, $\PXPbar=0$. The scalar contraction vanishes exactly, just as expected for massless particles. 
The full tensor collapses to a single rank-one, traceless object,
\begin{equation}
		t^{\mu\nu}=\frac{2h_S\,}{\tau \,|\pp|^{3}}\;P^{\mu}P^{\nu},
		\qquad
		P^{\mu}=\Big(\tfrac{|\pp|}{\sqrt{2}} \rme^\eta,\;\tfrac{|\pp|}{\sqrt{2}}\rme^{-\eta},\;\pp\Big)
\end{equation}
i.e. $P^\mu=p^\mu\big|_{y=\eta}$ is the on-shell massless gluon momentum evaluated at
$y=\eta$. In this limit $t^{\mu\nu}$ is exactly of Cooper--Frye form, with the hard factor
$h_S\,=S\bar S+A\bar A$ of \cref{eq:hard-factor-contracted}; the residual $\mathcal D$ terms
survive only at $\Pp\neq\bar{\Pp}$, i.e. as gradient corrections.


\bibliographystyle{JHEP}
\bibliography{References}

\end{document}